\documentclass[12pt]{article}
\usepackage{fancyhdr}
\usepackage{latexsym} 
\usepackage{amssymb}
\usepackage{graphicx}
\usepackage{epsf}
\usepackage{amscd}
\usepackage{amsmath}
\usepackage{color}

\makeatletter
\@addtoreset{equation}{section}

\renewcommand{\thefootnote}{\fnsymbol{footnote}}
\makeatother

\def\ee{\end{eqnarray}}

\newcommand{\nn}{\nonumber}
\def\p{\partial}

\def\D{\mathcal{D}}

\def\=:{=\hspace{-.7em}\raisebox{1.1ex}{.}\hspace{.1em}\raisebox{-0.2ex}{.} }

\newcommand {\beq}{\begin{eqnarray}}
\newcommand {\eeq}{\end{eqnarray}}
\newcommand {\non}{\nonumber\\}

\newcommand {\1}[1]{\frac{1}{#1}}
\newcommand {\2}[1]{\frac{i}{#1}}
\newcommand {\thb}{\bar{\theta}}

\newcommand {\psb}{\bar{\psi}}

\newcommand {\sig}{\sigma}
\newcommand {\sigb}{\bar{\sigma}}

\newcommand {\del}{\partial}
\newcommand {\dagg}{^{\dagger}}

\newcommand {\lam}{\lambda}

\newcommand{\hs}[1]{\hspace{#1 mm}}
\newcommand{\al}{\alpha}
\newcommand{\be}{\beta}

\newcommand {\Nf}{N_{\rm F}} 
\newcommand {\Nc}{N_{\rm C}}

\begin{document}
\thispagestyle{empty}

\begin{flushright}
IFUP-TH/2008-04, \;
TIT/HEP--580, \;
DAMTP-2008-9\\
arXiv:{\tt yymm.nnnn} [hep-th], \;\;
February, 2008 \\
\end{flushright}

\begin{center}
{\Large \bf 
Domain Walls with Non-Abelian Clouds
} 
\\[10mm]

{\normalsize\bfseries
Minoru~Eto$^{a,b}$, 
Toshiaki~Fujimori$^c$, 
Muneto~Nitta$^d$,\\
Keisuke~Ohashi$^e$, 
and
 Norisuke~Sakai$^c$}
\footnotetext{
e-mail~addresses: \tt
minoru(at)df.unipi.it;
fujimori(at)th.phys.titech.ac.jp;\\
nitta(at)phys-h.keio.ac.jp;
K.Ohashi(at)damtp.cam.ac.uk;
nsakai(at)th.phys.titech.ac.jp
}

\vskip 1.5em
$^a$ {\it INFN, Sezione di Pisa,
Largo Pontecorvo, 3, Ed. C, 56127 Pisa, Italy
}
\\
$^b$ {\it Department of Physics, University of Pisa
Largo Pontecorvo, 3,   Ed. C,  56127 Pisa, Italy
}
\\
$^c$ {\it Department of Physics, Tokyo Institute of
Technology, Tokyo 152-8551, Japan
}
\\
$^d$ 
{\it Department of Physics, Keio University, Hiyoshi, Yokohama,
Kanagawa 223-8521, Japan
}
\\
$^e$
{\it Department of Applied Mathematics and Theoretical Physics, \\
University of Cambridge, CB3 0WA, UK}
\vspace{5mm}
%
%

{\bf Abstract}\\[5mm]
{\parbox{13cm}{\hspace{5mm}
Domain walls in $U(N)$ gauge theories, 
coupled to Higgs scalar fields with degenerate masses, 
are shown to possess normalizable 
non-Abelian Nambu-Goldstone(NG) modes, 
which we call non-Abelian clouds.  
We construct the moduli space metric 
and its K\"ahler potential of 
the effective field theory on 
the domain walls, by focusing on two models:  
a $U(1)$ gauge theory with several charged Higgs fields, 
and a $U(N)$ gauge theory with $2N$ Higgs fields 
in the fundamental representation.  
We find that non-Abelian clouds 
spread between two domain walls  
and that their rotation induces 
long-range repulsive force,  
in contrast to a $U(1)$ mode in 
models with fully non-degenerate masses 
which gives short-range force. 
We also construct a bound state of dyonic domain walls  
by introducing the imaginary part of the Higgs masses. 
In the latter model we find that 
when all walls coincide 
$SU(N)_{\rm L} \times SU(N)_{\rm R} \times U(1)$ 
symmetry is broken down to $SU(N)_{\rm V}$,  
and $U(N)_{\rm A}$ NG modes and 
the same number of quasi-NG modes 
are localized on the wall. 
When $n$ walls separate, 
off diagonal elements of 
$U(n)$ NG modes have wave functions spreading 
between two separated walls (non-Abelian clouds),
whereas some quasi-NG modes turn to NG bosons 
as a result of further symmetry breaking 
$U(n)_{\rm V} \to U(1)_{\rm V}^n$. 
In the case of $4+1$ dimensional bulk, 
we can dualize the effective theory 
to the supersymmetric 
Freedman-Townsend model of 
non-Abelian 2-form fields.

}}
\end{center}
\vfill
\newpage
\setcounter{page}{1}
\setcounter{footnote}{0}
\renewcommand{\thefootnote}{\arabic{footnote}}

\section{Introduction}\label{INTRO}

The moduli space of solitons 
provides elegant description of their classical and quantum 
dynamics \cite{Manton:1981mp}. 
If a global symmetry of the theory is spontaneously broken 
by the presence of solitons, 
a part of the moduli space 
is parametrized by Nambu-Goldstone (NG) modes 
associated with that broken symmetry. 
The broken symmetry acts on the moduli space metric as an 
isometry, which sometimes makes it an interesting object 
and is useful to determine the metric.  
In the case of symmetry spontaneously broken in vacua, 
the low energy effective action 
of corresponding NG modes can be constructed 
from only the information of symmetry breaking pattern, 
by using the nonlinear realization method \cite{Coleman:1969sm}. 
In some cases the moduli space metric of solitons 
can be determined thoroughly by symmetry alone.
For instance in the case of Yang-Mills instantons,  
a single instanton solution in $SU(N)$ gauge theory 
can be obtained by embedding 
the minimal solution $A_{\mu}^{\rm BPST}$ 
of $SU(2)$ gauge theory
found by Belavin et al.~(BPST) 
with the position $x_0$ 
and the size $\rho$ 
\cite{Belavin:1975fg} 
as 
\beq
 && 
 A_{\mu} = U \left(  
  \begin{array}{ccccccc}
    A_{\mu}^{\rm BPST} (x_0,\rho) & 0 \cr
                                0 & ~~~~\,{\bf 0}_{N-2}
  \end{array}\right)
   U\dagg   \non
 && U \in {SU(N) \over SU(N-2) \times U(1)}. 
  \label{eq:instanton}
\eeq 
Here $U$ brings a solution to another solution with 
degenerate masses or tension, 
so it gives a coset space of the NG modes. 
The moduli space in this case 
can be written as ${\cal I}^{k=1}_N \simeq 
{\bf C}^2 \times {\bf R}^+ \times 
{SU(N) \over SU(N-2) \times U(1)}$ with 
${\bf C}^2$ and ${\bf R}^+$ parametrized by $x_0$ 
and $\rho$, respectively \cite{Dorey:2002ik}.\footnote{
As a result the moduli space ${\cal I}^{k=1}_N$ 
is a cone over a tri-Sasakian manifold.
}
The cone singularity of this moduli space 
correspond to a small instanton configuration, 
and is blown up in the case of the non-commutative 
${\bf R}^4$ with the noncommutativity parameter $\theta$ 
\cite{Nekrasov:1998ss}, 
to yield
\beq 
 {\cal I}^{k=1}_{N,\theta} \simeq 
 {\bf C}^2 \times T^* {\bf C}P^{N-1}. 
 \label{eq:single-instanton-moduli} 
\eeq
The moduli space of separated multiple instantons is 
a symmetric product of ${\cal I}^{k=1}_N$'s 
(or ${\cal I}^{k=1}_{N,\theta}$'s). 
The orbifold singularities of it are resolved by the 
Hilbert scheme resulting in the full moduli space 
(which is smooth for ${\cal I}^{k=1}_{N,\theta}$ 
but still contains small instanton singularities for 
${\cal I}^{k=1}_N$.) 

Similar structure has been recently found 
in the case of vortices in certain non-Abelian gauge theory 
\cite{Hanany:2003hp,Auzzi:2003fs}. 
A $U(N)$ gauge theory with $N$ Higgs fields in 
the fundamental representation, denoted by  
an $N$ by $N$ matrix $H$, admits 
a minimal vortex solution 
\beq
 && 
 H = U \left(  
  \begin{array}{ccccccc}
    H^{\rm ANO} (z-z_0) & 0 \cr
                  0 & \sqrt c {\bf 1}_{N-1}
  \end{array}\right)  U\dagg  , \hs{10}
F_{12} 
 = U \left(  
  \begin{array}{ccccccc}
    F_{12}^{\rm ANO} (z-z_0) & 0 \cr
                           0 & ~~~~\,{\bf 0}_{N-1}
  \end{array}\right)  U\dagg  ,  \non
 && U \in {SU(N) \over SU(N-1) \times U(1) } 
\simeq {\bf C}P^{N-1}
  \label{eq:vortex}
\eeq 
where the Abrikosov-Nielsen-Olesen 
\cite{Abrikosov:1956sx} vortex solution 
$(H^{\rm ANO},F_{12}^{\rm ANO})$ in the Abelian-Higgs model 
is embedded into the 
upper-most and left-most components 
of the $N$ by $N$ matrices of 
the Higgs fields $H$ and 
the gauge (magnetic) fields $F_{12}$ in the $x^1$-$x^2$ plane. 
Here $z=x^1+ix^2$ is a codimensional coordinate of vortices.
A remarkable point 
is that the vortex solution (\ref{eq:vortex}) 
contains 
non-Abelian orientational moduli ${\bf C}P^{N-1}$ 
as in 
the instanton solution (\ref{eq:instanton}), 
in addition to the translational moduli $z_0 \in {\bf C}$;  
The moduli space is \cite{Hanany:2003hp,Auzzi:2003fs}
\beq 
 {\cal V}^{k=1}_N \simeq {\bf C} \times {\bf C}P^{N-1}.
 \label{eq:single-vortex-moduli}
\eeq
Again the moduli space of separate multiple vortices 
is a symmetric product of ${\cal V}^{k=1}_N$'s. 
The full moduli space was constructed \cite{Eto:2005yh} 
in which it turns out to be smooth 
with resolving orbifold singularities similarly to instantons. 
These vortices are called ``non-Abelian vortices" because 
the unbroken symmetry of the vacuum is non-Abelian. 
In general, when solitons exist in a 
symmetry breaking $G \to H$ with non-Abelian group $H$, 
they are called non-Abelian solitons 
irrespective of whether $H$ is a gauge (local) 
or global symmetry.\footnote{
In the case of instantons $H$ is $G$ itself because 
the symmetry $G$ is not broken.
} 
Then non-Abelian solitons are usually accompanied by
non-Abelian orientational moduli. 
See \cite{Tong:2005un,Eto:2006pg,Shifman:2007ce} for a review. 
It was observed by Hanany and Tong \cite{Hanany:2003hp} that 
the moduli space of non-Abelian vortices 
is a certain middle-dimensional submanifold 
of the moduli space of non-commutative instantons. 
In fact, the moduli space ${\cal V}^{k=1}_{N}$ 
in (\ref{eq:single-vortex-moduli})
of the single vortex solution (\ref{eq:vortex}) is 
a special Lagrangian submanifold of the moduli space 
${\cal I}^{k=1}_{N,\theta}$ in (\ref{eq:single-instanton-moduli})
of the single non-commutative instanton.
Physically this correspondence may be understood 
by the fact that instantons become vortices 
(sigma model instantons)
if they lie inside a vortex \cite{Hanany:2004ea,Eto:2004rz}.

Similarly to instantons and vortices, 
a correspondence between 
``Abelian" monopoles and ``Abelian" domain walls was found 
by Hanany and Tong \cite{Hanany:2005bq}.  
The 't Hooft-Polyakov monopoles \cite{'t Hooft:1974qc} 
are called Abelian 
because they occur when a gauge symmetry $G$ is broken 
to an Abelian subgroup $H$ of $G$. 
Typically it is $G=SU(2) \to H=U(1)$.  
In this case each monopole carries the moduli 
\beq 
 {\cal M}^{k=1} \simeq {\bf R}^3 \times S^1
 \label{eq:single-monopole}
\eeq 
where ${\bf R}^3$ corresponds to the position 
and $S^1$ to the phase of the internal space. 

In this paper we study domain walls in supersymmetric 
gauge theories (and corresponding nonlinear sigma models) 
with eight supercharges. 
So far, domain walls with eight supercharges 
have been mostly considered in 
gauge theories with
$U(1)$ gauge field \cite{Abraham:1992vb}--\cite{Sakai:2005kz}  
or $U(N)$ gauge fields \cite{Isozumi:2004jc}--\cite{Eto:2005wf} 
coupled to Higgs scalar fields with 
{\it non-degenerate} masses 
except for \cite{Shifman:2003uh,Eto:2005cc}. 
In the case of non-degenerate Higgs masses, 
the flavor symmetry is Abelian: $U(1)^{N_{\rm F}-1}$ 
and the symmetry of the vacua is also Abelian. 
As a result each domain wall carries a $U(1)$ 
orientational modulus \cite{Abraham:1992vb,Isozumi:2004jc};
The moduli space of single domain wall is 
\beq
 {\cal W}^{k=1} \simeq {\bf R} \times S^1. \label{eq:mod-wall}
\eeq 
From this viewpoint,  
these domain walls should be 
called Abelian domain walls 
even when  the gauge symmetry of the Lagrangian 
is non-Abelian \cite{Isozumi:2004jc}--\cite{Eto:2005wf}.\footnote{
In our early papers 
\cite{Isozumi:2004jc}--\cite{Eto:2005wf} 
we called these solutions non-Abelian domain walls 
because of the non-Abelian gauge symmetry,  
but this is not appropriate 
in the current definition of 
non-Abelian solitons.
}

The moduli space (\ref{eq:mod-wall})
of a single domain wall  
is a middle dimensional submanifold of 
the moduli space (\ref{eq:single-monopole}) of 
a single Abelian monopole as discussed in \cite{Hanany:2005bq}.
The moduli space of multiple domain walls 
was constructed in $U(N_{\rm C})$ gauge theory 
coupled to Higgs fields
with non-degenerate masses \cite{Isozumi:2004jc,Isozumi:2004va}, 
and the correspondence to the multi-monopole moduli space was studied 
in \cite{Hanany:2005bq} as  
noted  
above. 
Similarly to the correspondence between instantons and vortices, 
this correspondence may be understood by 
noting that 
monopoles become domain walls inside a vortex 
\cite{Tong:2003pz,Hanany:2004ea,Shifman:2004dr}.

Non-Abelian monopoles appear when gauge symmetry 
$G$ is broken down to non-Abelian subgroup $H$ 
\cite{NA-monopoles}--\cite{Auzzi:2004if}
which is the case that some vacuum expectation values (VEV) of 
adjoint Higgs fields are degenerate. 
As a result non-Abelian zero modes appear 
around the non-Abelian monopoles.
Some of these zero modes are normalizable modes 
which are easy to deal with;  
When we turn on a small difference in VEVs 
of adjoint Higgs fields with degenerate VEVs, 
one non-Abelian monopole is split into 
two Abelian monopoles, 
the light monopole with the mass  
corresponding to the small difference 
between the VEVs 
and the one with almost the same 
mass of the original non-Abelian monopole. 
When 
the difference between the VEVs decreases, 
the light monopole grows 
with 
the size bounded from the above by 
the distance to the other monopole. 
This mode was called the 
{\it non-Abelian cloud} by Eric Weinberg 
\cite{NA-monopoles}. 
However the other modes around non-Abelian monopoles 
are non-normalizable   
and cannot be considered as moduli of monopoles themselves. 
The latter makes the study of non-Abelian monopoles difficult,  
which is in fact a notorious problem.
Non-Abelian monopoles are important ingredients for 
a non-Abelian extension of duality in 
supersymmetric gauge theories \cite{Dorey:1996jh,Auzzi:2004if}.

Our concern in this paper is about 
domain walls with 
non-Abelian orientational moduli, 
which may be called {\it non-Abelian domain walls}.   
One expects that the relation 
found by Hanany-Tong \cite{Hanany:2005bq} between 
Abelian monopoles and Abelian domain walls 
can be extended to the one 
between {\it non-Abelian} monopoles and 
{\it non-Abelian} domain walls.  
One motivation to study non-Abelian domain walls 
is to obtain a hint to understand non-Abelian monopoles 
through this correspondence.  
Unlike instantons or vortices, 
Higgs fields need masses for domain walls to exist. 
Once the Higgs masses are (partially) {\it degenerate}, 
the model exhibits a non-Abelian flavor symmetry $G$
and the vacua break $G$ into its non-Abelian subgroup $H$.
Then the domain wall solutions further break the non-Abelian symmetry 
$H$ of vacua and 
are expected to acquire 
non-Abelian orientational moduli 
associated with the breaking of $H$, 
resulting in  
non-Abelian domain walls. 
In fact $U(2)$ moduli were already found 
by Shifman and Yung 
\cite{Shifman:2003uh,Eto:2005cc} 
in the $U(2)$ gauge theory coupled to four charged 
Higgs fields with 
the common $U(1)$ charge and 
the mass matrix $M = {\rm diag} (m,m,-m,-m)$.

In this paper we study zero modes 
of non-Abelian domain walls and their properties 
in two different models.
The first model is a $U(1)$ gauge theory with 
$N_{\rm F}$ Higgs fields with an $N_{\rm F}$ by $N_{\rm F}$ 
mass matrix $M= {\rm diag} (m_1,0,\cdots,0,-m_2)$. 
The second model is 
a $U(N)$ gauge theory with 
$N_{\rm F} = 2N$ Higgs fields in the fundamental representation, 
with the half of the Higgs masses being $-m$ and the rest being $m$.   
We call the latter the generalized Shifman-Yung (GSY) model 
because the case of $N=2$ was discussed by Shifman and Yung
\cite{Shifman:2003uh,Eto:2005cc}.
We construct the K\"ahler potential and the metric of 
the effective Lagrangian 
of normalizable zero modes (moduli)
of domain walls in these two models,
by using the recently developed method to 
obtain effective Lagrangian on the BPS solitons \cite{Eto:2006uw}. 
It is a supersymmetric nonlinear sigma model with 
the moduli space of domain walls as 
its target space. 
We find that the target space of the first model is 
${\bf C}^* \times {\bf C}^{N_{\rm F}-2}$,  
equipped with a non-flat metric for the latter,  
on which the isometry ${\bf C}^* \times U(N_{\rm F}-2)$ acts. 
The target space of the second (GSY) model  
turns out to be  
$GL(N,{\bf C}) \simeq {\bf C}^* \times SL(N,{\bf C})$ 
on which the isometry 
${\bf C}^* \times SU(N)_{\rm L} \times SU(N)_{\rm R}$ acts.
We find 
the following. 
When positions of all domain walls coincide,  
$SU(N)_{\rm V}$ symmetry is preserved 
and the massless Nambu-Goldstone modes 
$[SU(N)_{\rm L} \times SU(N)_{\rm R}] / SU(N)_{\rm V} \simeq
SU(N)_{\rm A}$,
associated with 
the non-Abelian flavor symmetry breaking 
$SU(N)_{\rm L} \times SU(N)_{\rm R} \to SU(N)_{\rm V}$, 
are localized at the coincident wall.  
When $n$ (among $N$) domain walls are separated, however, 
the $SU(n)_{\rm V}$ subgroup of $SU(N)_{\rm V}$ 
is further broken down to $U(1)_{\rm V}^{n-1}$. 
Consequently 
only the diagonal $U(1)_{\rm A}^{n-1}$ Nambu-Goldstone modes 
in $SU(n)_{\rm A} [\subset SU(N)_{\rm A}]$ are localized 
on each individual wall and 
the off-diagonal Nambu-Goldstone modes in $SU(n)_{\rm A}$   
have wave functions spreading between a set of two 
separated walls. 
The latter can be called {\it non-Abelian clouds} 
because corresponding modes have been 
introduced in the context of non-Abelian monopoles 
\cite{NA-monopoles}. 
We find that 
these non-Abelian clouds remain massless 
in the GSY model 
when domain walls are separated.\footnote{
In reference \cite{Shifman:2003uh}, the authors 
argued that these modes spreading between walls become 
massive, contrary to our results.  
}

In the above we see that the number of NG modes 
can change depending on the positions of the walls. 
A question is whether the number of massless modes  
or dimensionality of the moduli space changes or not. 
The answer is no; the total number of massless modes is preserved. 
Key ingredients to understand this phenomenon are 
so-called quasi-Nambu-Goldstone modes 
which do not directly correspond to
underlying spontaneously broken global symmetry 
but are required from unbroken supersymmetry 
\cite{Bando:1983ab,Kotcheff:1988ji}.\footnote{ 
These massless bosons are considered in the context 
of the preon models and 
the nonlinear realization of spontaneously broken global symmetries  
with preserving supersymmetry. 
The additional massless fermions to constitute the chiral multiplets 
are called quasi-Nambu-Goldstone fermions \cite{Buchmuller:1982xn}. 
The presence of these massless 
non-Abelian clouds 
is a distinguishing feature of the walls with non-Abelian flavor 
symmetry at the classical level, in contrast to 
the open string modes becoming massive when D-branes are 
separated. 
}  
When all the domain walls coincide 
there exist quasi-NG modes as many as $SU(N)_{\rm A}$ NG modes. 
Among them diagonal $N-1$ modes represent positions of the walls. 
When $n$ walls separate,  
some quasi-NG modes turn to 
the NG modes $SU(n)_{\rm V}/U(1)_{\rm V}^{n-1}$ 
for the further symmetry breaking 
 $SU(n)_{\rm V} \to U(1)_{\rm V}^{n-1}$.
Therefore the quasi-NG modes and NG modes can change 
to each other 
with the total number of massless modes unchanged. 
All of these states with different 
symmetry breaking patterns are degenerate,  
which was originally found by G.~Shore \cite{Kotcheff:1988ji} 
in the context of supersymmetric nonlinear realizations.  

We also construct the Lagrangian 
in a dual description by 2-form fields 
on the domain wall world volume,  
when domain walls (with 3+1 dimensional world volume) 
exist in $d=4+1$ dimensions. 
This is in contrast to the $2+1$ dimensional world-volume, 
where vector fields 
in a dual description have been obtained only for free field part 
without interactions \cite{Shifman:2003uh}. 
In the case of the GSY model, 
we can obtain 
the supersymmetric extension \cite{Clark:1988gx} 
of the so-called Freedman-Townsend model 
\cite{Freedman:1980us} of 
non-Abelian 2-forms with non-trivial interaction.

Although we have emphasized the importance 
of the relation to non-Abelian monopoles in this introduction, 
this work may have some impacts on the brane-world scenario 
\cite{brane-world} too. 
Our model can be made in dimension $d=4+1$ so that 
we have domain walls as branes with $3+1$ dimensional world volume 
and ${\cal N}=1$ supersymmetry. 
The non-Abelian clouds found in this paper 
have a wave function spreading between two branes. 
One brane has an interaction from another brane 
mediated by these inter-brane modes.

This paper is organized as follows.
In Sec.~\ref{sc:NAWall-deg-mass}, 
we define our model and review 
methods to obtain the effective Lagrangian on walls. 
In Subsec.~\ref{subsec:local}
the local structure of the moduli space is investigated.
In Subsec.~\ref{subsec:D-brane}
we understand it by means of the kinky D-brane configurations.

In Sec.~\ref{section:3} we study 
the $U(1)$ gauge theory with $N_{\rm F}$ charged Higgs scalar fields
as the simplest model 
of domain walls with degenerate Higgs masses. 
First of all in Subsec.~\ref{sc:simplest} 
we study the simplest case of $N_{\rm F}=4$ 
to show the behavior of non-Abelian clouds.
To this end we introduce a small mass spiting $\epsilon$ 
in degenerate masses so that 
we have a domain wall with 
a small tension (proportional to $\epsilon$) 
between the two usual domain walls. 
As the mass splitting decreases $\epsilon \to 0$, 
the width of such a domain wall 
(proportional to $\epsilon^{-1}$) grows. 
In the end the domain wall profile is 
bounded by the positions 
of the neighboring two domain walls 
and fills between them. 
Thus it 
becomes a non-Abelian cloud.  
This technique was used by E. Weinberg to study non-Abelian clouds 
in non-Abelian monopoles 
\cite{NA-monopoles}. 
In Subsec.~\ref{subsec:efac} we construct the K\"ahler potential 
of the effective action for the moduli of 
the domain walls with general $N_{\rm F}$.  
The moduli space is 
${\bf C}^* \times {\bf C}^{N_{\rm F}-2}$ 
with the isometry ${\bf C}^* \times U(N_{\rm F}-2)$. 
In order to study the dynamics of the domain walls, 
we construct the conserved charges of 
the isometry $U(2)$ in the case of $N_{\rm F}=4$ for 
degenerate masses.  
We find that the two kinds of repulsive forces exist 
between the two domain walls with distance $R$; 
one comes from the $U(1)$ part of the isometry 
and its potential exponentially approaches to a constant 
as the distance $R$ becomes large, and 
the other comes from 
the $SU(2)$ part of the isometry 
and its potential behaves as $1/R$. 
The former has been known in the model with 
non-degenerate masses, which is mediated by massive modes 
between the two walls. 
The latter is new and is mediated by the non-Abelian clouds 
which are massless modes propagating between the two walls. 
In Subsec.~\ref{subsec:kink-bound} 
we construct a bound state of domain walls 
by introducing additional masses in the imaginary parts 
of the Higgs fields. 
The additional masses introduce 
an attraction between 
the walls and then balance 
with the repulsion by 
the charges of the non-Abelian clouds. 
The bound state is a dyonic domain wall of a new kind. 

In Sec.~\ref{sc:GSY} we work out the generalized 
Shifman-Yung (GSY) model.  
After presenting the vacua in Subsec.~\ref{sc:strong-coupling}
we construct domain wall solutions in Subsec.~\ref{subsec:Gen-sol}. 
In Subsec.~\ref{subsec:sym} we study the symmetry structure of 
the moduli space of the domain walls in the GSY model. 
If the positions of all the domain walls coincide, 
$SU(N)_{\rm L} \times SU(N)_{\rm R} [\times U(1)_{\rm A}]$ 
is spontaneously 
broken to $SU(N)_{\rm V}$ in the presence of the domain walls.  
A part of moduli space is parametrized by the Nambu-Goldstone modes 
(we may call them pions in analogy with the chiral symmetry 
breaking in hadron physics) 
associated with this breaking. 
The rest is parametrized by quasi-Nambu-Goldstone modes 
which are required by unbroken supersymmetry.
Some of them correspond to the positions of domain walls. 
When the walls are separated, the symmetry $SU(N)_{\rm V}$ 
is further broken down to its subgroup 
and hence 
there are more Nambu-Goldstone modes. 
These Nambu-Goldstone modes at finite wall separation 
become the quasi-Nambu-Goldstone modes in the limit of 
coincident walls. 
In Subsec.~\ref{subsec:efac-GSY}
we construct the K\"ahler potential of the effective 
Lagrangian of domain walls 
with arbitrary gauge coupling constant.
The moduli space turns out to be 
$GL(N,{\bf C}) \simeq {\bf C}^* \times SL(N,{\bf C}) 
[\simeq U(N)^{\bf C} \simeq T^* U(N)]$ 
on which $SU(N)_{\rm L} \times SU(N)_{\rm R} \times U(1)_{\rm A}$ 
acts as the isometry. 
In Subsec.~\ref{subsec:localization} 
we study wave functions of the modes 
by taking 
the strong gauge coupling limit. 
We find that non-Abelian clouds are spread between domain walls. 
Finally in Subsec.~\ref{subsec:dynamics-GSY} 
we expand the effective Lagrangian around 
the configurations in the cases of 
coincident walls and 
well-separated walls. 
In the former case 
we obtain the chiral Lagrangian as expected. 
We then study the effect of imaginary masses of the Higgs fields, 
to obtain a pion mass term.

In Sec.~\ref{sc:duality} 
the duality transformation is  
performed for the massless particles to obtain the non-Abelian 
tensor multiplets. 

In Sec.~\ref{sc:monopole} we apply our results to 
non-Abelian monopoles confined by non-Abelian vortices 
in the Higgs phase. 
We briefly discuss a monopole-monopole bound state.

In Sec.~\ref{sc:conclusion} is devoted to 
conclusion and discussion.

\section{
Non-Abelian Walls with Degenerate Higgs Masses 
}
\label{sc:NAWall-deg-mass}

\subsection{Models, symmetry and vacua}\label{subsec:models}

We consider $U(N_{\rm C})$ gauge theory in 
space-time dimension 
from $d=1+1$ to $d=4+1$ with (at least one) 
real scalar field $\Sigma$ in the 
 adjoint representation and $N_{\rm F}\,(>N_{\rm C})$ 
flavors of 
massive Higgs 
scalar fields in the fundamental representation, 
denoted as an 
$N_{\rm C} \times N_{\rm F}$ matrix $H$. 
Choosing the minimal kinetic term, we obtain 
\begin{eqnarray}
{\cal L} &=& {\cal L}_{\rm kin} - V, 
\label{eq:mdl:total_lagrangian}
\\ 
{\cal L}_{\rm kin} &=& 
{\rm Tr}\left(- {1\over 2g^2}F_{\mu\nu}F^{\mu\nu} 
+\frac{1}{g^2}{\cal D}_\mu \Sigma \, {\cal D}^\mu \Sigma 
+{\cal D}^\mu H \left({\cal D}_\mu H\right)^\dagger \right), 
\label{eq:mdl:lagrangian}
\end{eqnarray}
where the covariant derivatives and field strengths are 
defined as 
${\cal D}_\mu \Sigma=\partial_\mu\Sigma + i[W_\mu, \Sigma]$, \hs{0.1}
${\cal D}_\mu H=(\partial_\mu + iW_\mu)H$, \hs{0.1}
$F_{\mu\nu}=-i[{\cal D}_\mu,\,{\cal D}_\nu]$. 
Our convention for the space-time metric is 
$\eta_{\mu\nu} = {\rm diag}(+,-,\cdots,-)$. 
The scalar potential $V$ is given in terms of a 
diagonal mass matrix $M$ and a real parameter $c$ as 
\begin{eqnarray}
V&=& 
{\rm Tr}
\Big[
\frac{g^2}{4}
\left(c\mathbf{1}
-H  H^{\dagger} 
\right)^2 
+ (\Sigma H - H M) 
 (\Sigma H - H M)^\dagger 
\Big] . 
\label{eq:mdl:scalar_pot}
\end{eqnarray}
This Lagrangian can be made supersymmetric 
by adding another scalar in the fundamental representation 
($H^1\equiv H$, $H^2=0$), and fermionic partners of all 
these bosons. 
The resulting theory has eight supercharges. 
We have chosen for simplicity the gauge couplings for 
$U(1)$ and $SU(N_{\rm C})$ to be identical to 
obtain simple solutions classically, even though 
they are independent. 
The real positive parameter $c$ 
is called the Fayet-Iliopoulos (FI) parameter, 
which can appear in supersymmetric $U(1)$ gauge theories 
\cite{Fayet:1974jb}.

Next let us discuss the vacuum structure of this model. 
In the case of massless Higgs fields ($M=0$), 
the Lagrangian enjoys 
a flavor symmetry $SU(\Nf)$ (the overall $U(1)$ is gauged in this model). 
The vacua constitute the Higgs branch, 
which is isomorphic to 
a hyper-K\"ahler manifold $T^* G_{\Nf,\Nc}$, 
the cotangent bundle over the Grassmann manifold \cite{Lindstrom:1983rt}
\beq 
  G_{\Nf,\Nc} \simeq {SU(\Nf) \over SU(\Nc) \times SU(\Nf-\Nc) \times U(1)}.
  \label{eq:massless-Higgs-br}
\eeq
The coset structure reflects the fact that 
the global symmetry $SU(\Nf)$ is spontaneously broken and 
that the Nambu-Goldstone bosons for the broken symmetry appear.

When masses are fully non-degenerate 
the flavor symmetry is explicitly broken down to 
$U(1)^{\Nf-1}$ 
and the vacua reduce to a finite number of discrete points \cite{Arai:2003tc}. 
The number of vacua is given by 
${\Nf ! \over \Nc ! (\Nf - \Nc)!}$. 
All the Nambu-Goldstone bosons become pseudo Nambu-Goldstone bosons 
with masses.
Domain wall solutions interpolating between these vacua 
were discussed so far \cite{Isozumi:2004jc}--\cite{Eto:2005wf}.

On the other hand, 
when the Higgs masses are partially degenerate 
the flavor symmetry is enhanced as
\beq 
 U(1)^{\Nf-1} \rightarrow SU(N_1) \times SU(N_2) \times \cdots 
  \label{eq:flavor}
\eeq
with $N_i$ masses are degenerate ($i=1,2,\cdots$). 
There appear 
Nambu-Goldstone modes continuously parametrizing degenerate vacua, 
which constitute a submanifold of 
the massless Higgs branch $T^* G_{\Nf,\Nc}$.
This is the situation which we 
consider in this paper.

\subsection{BPS equations and the moduli matrix}\label{subsec:BPSeq}

The 1/2 BPS equations for domain walls 
interpolating the discrete vacua can be obtained 
by usual Bogomol'nyi completion of the energy 
\beq
E &=& \int_{-\infty}^{\infty} dy \, 
{\rm Tr} \left[ (\mathcal D_y H - HM + \Sigma H)^2 
+ \frac{1}{g^2} \left( \D_y \Sigma - \frac{g^2}{2} 
\left( c {\bf 1} - HH^\dagger\right) \right)^2 
+ c \, \D_y \Sigma \right] \notag \\
&\geq& c \Big[{\rm Tr} \, \Sigma(\infty) 
- {\rm Tr} \, \Sigma(-\infty) \Big].
\eeq
The first order differential equations for the 
configurations saturating this energy bound are of the 
form \cite{Isozumi:2004jc} 
\beq
\mathcal D_y H = HM - \Sigma H, \qquad \mathcal D_y \Sigma
= \frac{g^2}{2} \left(c{\bf 1} - HH^\dagger \right).
\label{eq:BPS}
\eeq
Here we consider static configurations 
depending only on the $y$-direction.

Let us solve these 1/2 BPS equations.
Firstly the first equation can be solved by \cite{Isozumi:2004jc}
\begin{align}
H=S^{-1}(y)H_0e^{My},\quad 
\Sigma+iW_y=S^{-1}(y)\partial_y S(y).
\label{eq:solBPS}
\end{align}
Here $H_0$, 
called the moduli matrix, 
is an $\Nc \times \Nf$ 
constant complex matrix of rank $\Nc$, 
and contains all the moduli parameters of solutions. 
The matrix valued quantity $S(y) \in GL(\Nc,{\bf C})$ is determined by 
the second equation in (\ref{eq:BPS}) which can be converted 
to the following equation for $\Omega \equiv SS^\dagger$:
\begin{equation}
\frac{1}{cg^2} \bigl[ \partial_y(\Omega^{-1}\partial_y \Omega) \bigl] =
{\bf 1}_{\Nc}-\Omega^{-1}\Omega_0,\hs{10}
\Omega_0 \equiv \frac{1}{c} H_0 e^{2My} H_0^\dag .
\label{eq:master}
\end{equation}
This equation is called the master equation for domain walls. 
From the vacuum conditions 
at spatial infinities $y \rightarrow \pm \infty$, 
we can see that the solution $\Omega$ of the master equation should 
satisfy the boundary condition $\Omega \rightarrow \Omega_0$ 
as $y \rightarrow \pm \infty$. 
It determines $S$ for a given moduli matrix $H_0$ 
up to the gauge transformations $S^{-1} \rightarrow U S^{-1}, ~ U \in U(\Nc)$ 
and then the physical fields can be obtained through (\ref{eq:solBPS}). 
Note that the master equation is symmetric under the following 
$V$-transformations
\begin{eqnarray}
H_0 \rightarrow VH_0 \quad 
{\rm and} \quad S(y) \rightarrow VS(y) \quad  {\rm with} \quad 
V \in GL(\Nc, {\bf C}), 
\label{eq:Vtransformation}
\end{eqnarray}
and if the moduli matrices are related by the $V$-transformations 
$H_0' = V H_0$, they give physically equivalent configurations. 
We call this equivalence relation as the 
$V$-equivalence relation and denote it as $H_0 \sim V H_0$. 
The master equation was shown to be non-integrable \cite{Inami:2006wr}, 
and the existence and uniqueness of its solution for 
any given $H_0$ was rigorously proved at least for 
the $U(1)$ gauge theory \cite{Sakai:2005kz}.

In the effective action on the domain walls, the moduli parameters $\phi^i$ appearing in the moduli matrix $H_0$ are promoted to fields $\phi^i(x^\mu)$ which depend on the coordinates of the world-volume. 
Then the effective theory is described as a nonlinear sigma model whose target space is the moduli space endowed with a K\"ahler metric. 
The K\"ahler metric of the effective action can be obtained through the K\"ahler potential which is written down as the following integral form 
\cite{Eto:2006pg,Eto:2006uw}
\begin{eqnarray}
K(\phi,\phi^*) &=& \int^{\infty}_{-\infty} dy 
 \, \Big[ {\cal K}(y,\phi,\phi^*) 
 - {\cal K}_{ct}(y,\phi) - \bar{\cal K}_{ct}(y,\phi^*) \Big],
  \label{eq:formal-Kahler} \\ 
{\cal K}(y,\phi,\phi^*) 
 &=&{\rm Tr}\left[c\,\log\Omega+c\,\Omega^{-1}\Omega_0 
  +\frac{1}{2g^2}\left(\Omega^{-1}\partial_y\Omega\right)^2\right],
\label{eq:formal-Kahler-density0}
\end{eqnarray}
where ${\cal K}_{ct}(y,\phi)$ and $\bar{\cal K}_{ct}(y,\phi^\ast)$ are counter terms, which are added to subtract the divergent part contained in ${\cal K}(y,\phi,\phi^\ast)$. Note that this addition of the counter terms can be interpreted as the K\"ahler transformation and the K\"ahler metric $K_{ij} = \frac{\p K}{\p \phi^i \p \phi^{j\ast}}$ does not change by the addition of the counter terms ${\cal K}_{ct}(y,\phi)$, ${\cal K}'_{ct}(y,\phi^\ast)$ which are purely holomorphic and anti-holomorphic with respect to the moduli parameters respectively.

According to this formula, as a matter of course we can confirm that
the total inertial mass of the walls, $T_{\rm inertial}$,
 agrees with the total static energy (tension) of the BPS walls, 
$T_{\rm BPS}$, by the following discussion.
Assume a field $w(x^\mu)$ consists of the center of masses, 
${\rm Re}\, w$, and a Nambu-Goldstone mode for the overall phase, 
${\rm Im}\, w$.  
The total inertial mass $T_{\rm inertial}$ 
is given by the coefficient of the kinetic
term of the center of mass, $2 K_{ww^*}$.
Because of the translational invariance, $w$ and the coordinate $y$ 
appear in the K\"ahler potential density ${\cal K}$
(\ref{eq:formal-Kahler-density0}) through a form, $y-{\rm Re}\,w$.
This fact leads to the statement above as, 
\begin{eqnarray}
T_{\rm inertial} &=& 2\frac{\partial^2 K}{\partial w \partial w^*} ~=~ \frac12 \int^{\infty}_{-\infty} dy \, \frac{\partial^2 {\cal K}}{\partial y^2} 
\nn\\
&=& \frac{c}{2} \left[\frac{\partial }{\partial y}{\rm Tr } \log \Omega 
\right]^{y=\infty}_{y=-\infty} = c \Big[{\rm Tr}\Sigma \Big]^{y=\infty}_{y=-\infty} = T_{\rm BPS}. \label{eq:inertial-mass}
\end{eqnarray}
For well-separated walls, this statement is also applicable to each wall
and determines an asymptotic metric for their position moduli.
Combining this and the flavor symmetry of the system,
we can often determine the asymptotic metric for full moduli space.   
This is a main strategy in Section 3 and Section 4.

The technique introduced here to solve BPS equations, 
the moduli matrix formalism, was generalized to non-Abelian 
vortices in various cases \cite{Eto:2006mz}; 
changing the manifold from ${\bf R}^2$ to a cylinder or 
a torus $T^2$, non-Abelian string reconnection, 
an extension to the semi-local case and the finite temperature. 
See Refs.~\cite{Eto:2006pg,proceedings} for a review 
including other composite BPS solitons.

\subsection{Domain walls and local structure of the moduli space}
\label{subsec:local}

We now discuss the domain wall solutions 
interpolating between different vacua. 
Domain walls in the case of the fully non-degenerate Higgs 
masses were constructed and discussed 
previously 
\cite{Abraham:1992vb}--\cite{Eto:2004vy}.
In the $U(1)$ gauge theory, 
the model admits the $N$ ordered vacua and 
the $N-1$ domain walls connecting them. 
Each wall carries 
a zero mode of broken $U(1)$ symmetry and 
a broken translational symmetry. 
Rigorously speaking, only one massless field is the exact 
Nambu-Goldstone mode for the broken translational 
symmetry. 
The others are approximate Nambu-Goldstone modes when 
all walls are far away from each other. 
Then each wall carries a 
zero mode locally in moduli space 
\beq 
 {\bf C}^* (= {\bf C} - \{0\}) \simeq {\bf R} \times U(1).  
\label{eq:wall-localized}
\eeq 
However one has to note that 
the moduli space of the full solution 
is {\it not} a direct product of them. 
For instance let us consider $U(1)$ gauge theory 
with three flavors. 
This model contains three isolated vacua and 
admits two walls. 
The moduli space of two domain walls 
is not a direct product of two ${\bf C}^*$'s 
but ${\bf C}^* \times {\bf C}$. 
This is because two walls cannot pass through, 
and one of $U(1)$ modulus shrinks 
when they are compressed to a single wall.

\begin{figure}[thb]
\begin{center}
\includegraphics[width=8cm,clip]{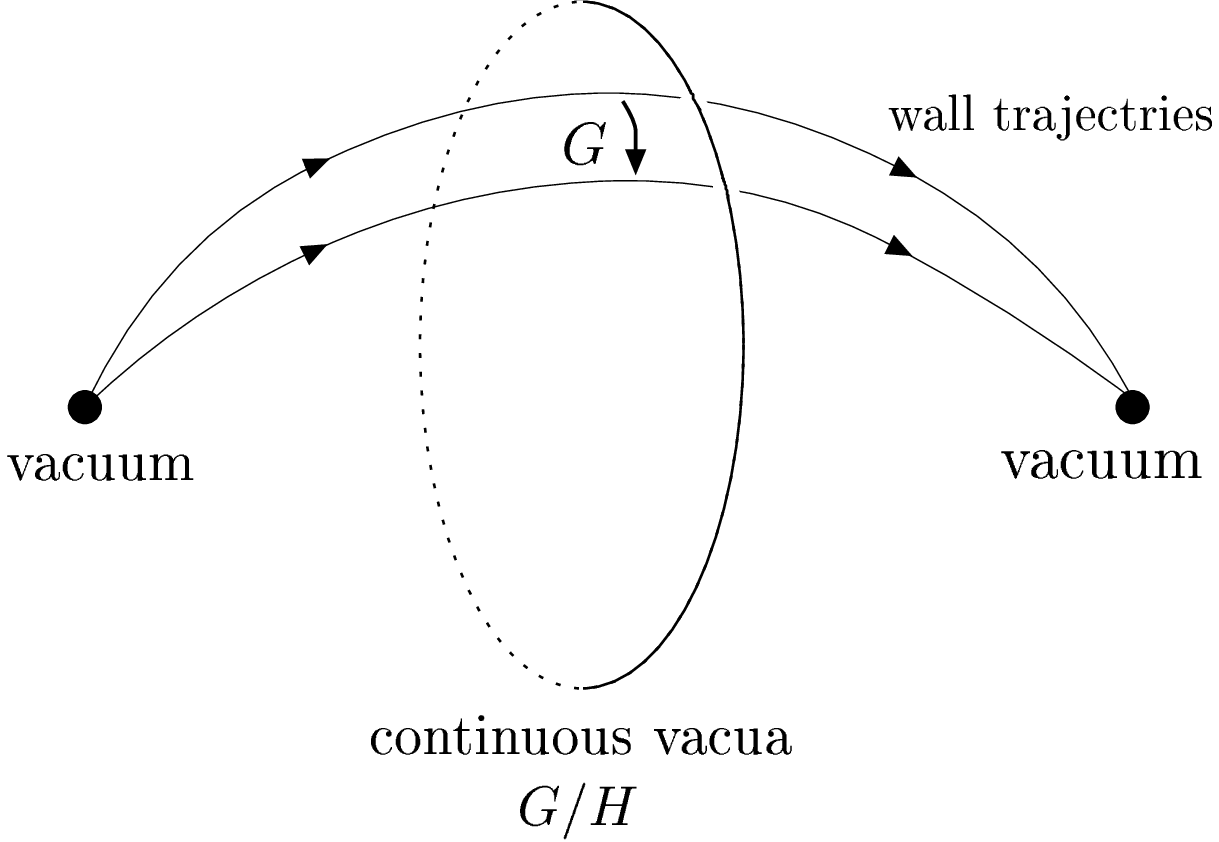}
\end{center}
\caption{\small 
Continuous series of wall solutions parametrized by 
$G/H$ are obtained when trajectories pass near the 
continuously degenerate vacua. 
}
\label{fig1}
\end{figure}

Continuously degenerate vacua occur 
when a global symmetry $G$ is spontaneously broken. 
If it breaks to its subgroup $H$, 
the Nambu-Goldstone bosons parametrizing a coset space 
$G/H$ appear. 
Let us consider the situation such 
that a path of a wall configuration, 
connecting two isolated vacua, 
passes near the continuously degenerate vacua.
Once a wall solution is found, another solution 
can be obtained by acting the global symmetry 
$G$ on it. 
Then we obtain a continuous series of solutions 
parametrized by $G/H$ 
as shown in fig.~\ref{fig1}. 
In other words, 
non-Abelian Nambu-Goldstone modes of $G/H$ 
are localized on the wall solution 
since $G$ fixes the two isolated vacua. 

Actually the condition that both vacua on both sides 
of the wall are isolated is not necessary. 
Rather this localized non-Abelian zero modes 
usually occur when a wall configuration 
passes near continuously degenerate vacua 
as seen in the next subsection.

\subsection{D-brane configurations}\label{subsec:D-brane}

The wall configurations are realized as
a kinky D-brane configuration \cite{Eto:2004vy}. 
(See \cite{Lambert:1999ix} for the case of $U(1)$ gauge group.)
In this subsection we generalize the discussion of \cite{Eto:2004vy} 
to the case of partially degenerate Higgs masses. 
We will see the D-brane configuration 
is very useful to understand a local structure of the moduli space 
of domain wall solutions.

\begin{figure}[thb]
\begin{center}
\includegraphics[width=8cm,clip]{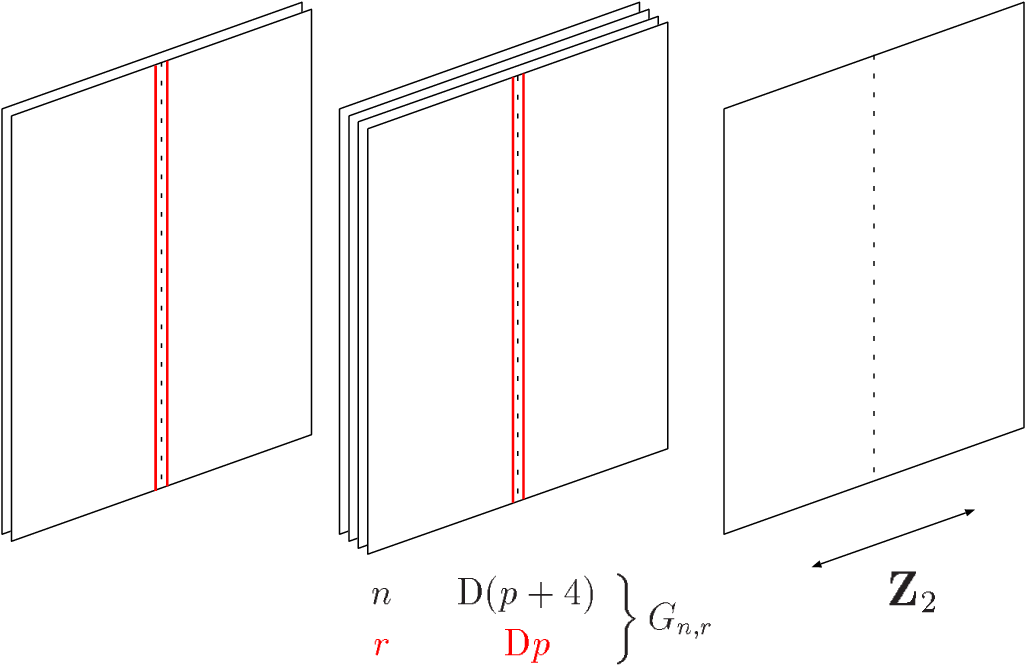}
\end{center}
\caption{\small
D-brane configurations for 
a degenerate vacuum $G_{n,r}$.
}
\label{fig2}
\end{figure}

First of all the model in $d=p+1$ dimensions ($p=1,2,3,4$)
can be realized on
a $p+1$ dimensional world-volume of D$p$-branes 
in a D$p$-D$(p+4)$ system. 
The four codimensional direction ${\bf C}^2$ of the 
D$p$-branes along D$(p+4)$-branes are divided by 
${\bf Z}_2$ in order to remove 
unwanted adjoint Higgs fields describing the positions 
of the D$p$-branes inside the D$(p+4)$-branes. 
Then we can regard D$p$-branes as fractional D($p+2$)-branes 
stacked at orbifold singularity of ${\bf C}^2/{\bf Z}_2$. 
(Taking T-duality we can map the brane configuration to 
a D($p+1$)-D($p+3$)-NS5 system of the Hanany-Witten set up, 
but we do not do that in this paper.) 
Hypermultiplets containing Higgs fields 
are obtained from strings connecting 
the D$p$- and D$(p+4)$-branes whereas 
vector multiplets containing gauge fields 
appear from strings connecting 
the D$p$-branes.  
When the positions of D($p+4$)-branes 
split along their codimensions in ten dimensions, 
the Higgs fields (the hypermultiplets) get masses. 
In order to discuss domain walls 
we consider here real masses which is allowed for any 
dimensions.\footnote{
We need complex Higgs masses when we 
construct domain wall junctions (network or webs) 
\cite{Eto:2005cp,Eto:2007uc}
or dyonic domain walls \cite{Lee:2005sv,Eto:2005sw}. 
The complex masses is possible up to four dimensions 
($p=1,2,3$). 
}
Previously we considered the fully non-generate masses 
and therefore completely separated D($p+4$)-branes 
\cite{Eto:2004vy}. 
Now we consider the case that 
$N_i \, (i=1,2,\cdots)$ coincident D($p+4$)-branes 
realizing the flavor symmetry (\ref{eq:flavor}). 
In a vacuum where each D$p$-brane sits in one of the 
D($p+4$)-branes, 
at most $n$ D$p$-branes can coexist in the 
$n$ coincident D($p+4$)-branes due to the requirement of 
the so-called s-rule \cite{HananyWitten97}. 
The vacuum configuration can be illustrated as Fig.~\ref{fig2}.

We can find the vacuum structure from this configuration. 
When $r$ D$p$-branes sit in $n (> r)$ coincident 
D($p+4$)-branes, we obtain degenerate vacua, 
the cotangent bundle $T^* G_{n,r}$  
over the Grassmann manifold 
(see Fig.~\ref{fig2}),
\beq
 G_{n,r} \simeq {SU(n) \over SU(r) \times 
SU(n-r) \times U(1)}. 
   \label{eq:Gnr}
\eeq
This manifold is a submanifold of the massless Higgs branch 
(\ref{eq:massless-Higgs-br}).
We thus find that the moduli space of vacua 
is the direct product of the Grassmann manifolds (\ref{eq:Gnr}): 
$\prod_{i}G_{n_i,r_i}$ ($0 \leq r_i \leq n_i$) with 
$\sum n_i = N_{\rm F}$ and 
$\sum r_i = N_{\rm C}$.

\begin{figure}[thb]
\begin{center}
\includegraphics[width=7cm,clip]{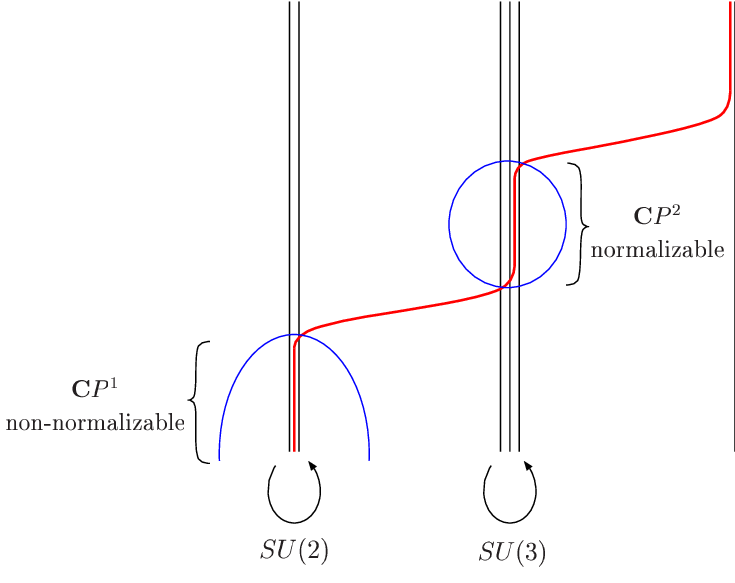}
\end{center}
\caption{\small
The D-brane configuration for 
two degenerate walls at the left, 
three degenerate walls in the middle, and a single wall 
at the right. 
}
\label{fig3}
\end{figure}

\begin{figure}[thb]
\begin{center}
\includegraphics[width=4cm,clip]{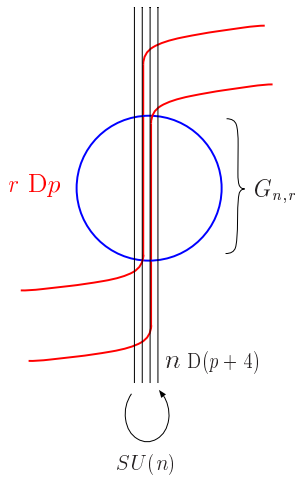}
\end{center}
\caption{\small
The $r(<n)$ D$p$-branes residing in finite region 
of $n$ D($p+4$) brane gives 
the zero modes forming the Grassmann manifold $G_{n,r}$. 
} 
\label{fig4}
\end{figure}

We now consider domain wall configurations. 
Eigenvalues of the adjoint Higgs field $\Sigma$ 
correspond to the positions of D$p$-branes. 
When there exists a domain wall, 
some (not necessarily one) D$p$-branes 
exhibit a kink, namely travel from one D($p+4$)-brane 
to another D($p+4$)-brane. 
The BPS condition dictates that these kinks  
have to move in one direction. 
An example of domain walls in a $U(1)$ gauge theory 
is drawn in Fig.~\ref{fig3}. 
From this configuration we can find 
zero modes associated with symmetry breaking. 
For instance on the middle D($p+4$)-branes in Fig.~\ref{fig3}
a D$p$-brane breaks $SU(3)$ flavor symmetry to 
$SU(2) \times U(1)$. 
Therefore associated with this symmetry breaking,  
there appear zero modes ${\bf C}P^2 \simeq SU(3)/[SU(2) \times U(1)]$. 
These modes are normalizable 
because this symmetry breaking occurs 
in a finite region between upper kink and the lower kink. 
This means that these modes have a support between 
the two domain walls. 
We call these modes as 
``non-Abelian clouds" as in the case of non-Abelian 
monopoles \cite{NA-monopoles}. 
In general when $r(<n)$ D$p$-branes exist at finite region 
of $n$ D($p+4$) brane there appear zero modes 
of the Grassmann manifold $G_{n,r}$ given in Eq.~(\ref{eq:Gnr}). 
See Fig.~\ref{fig4}. 
On the other hand, there also exist 
usual modes (\ref{eq:wall-localized}) localized on a wall which 
we call ``wall-localized modes". 

When a symmetry breaking  
occurs in an infinite or semi-infinite region 
as in the left-most part  
of Fig.~\ref{fig3}, the modes for this symmetry breaking 
have an infinite or semi-infinite 
support, and therefore they are non-normalizable. 
These bulk modes do not appear in the effective theory on walls, 
and do not contribute to the moduli space of walls.

In summary there in general 
appear normalized modes, classified into 
wall-localized modes and non-Abelian clouds, 
as well as non-normalizable modes. 
We can find a local structure of the moduli space 
but unfortunately at this stage 
we cannot find a global structure of
the moduli space from the brane configuration. 
In general
each part is not 
a direct product in the whole moduli space  
because of a non-trivial bundle structure. 
We have to integrate the modes over the codimension 
in order to obtain the whole moduli space.
We perform the integration 
explicitly in two examples in the succeeding sections.

\section{Non-Abelian Clouds 
in Abelian Gauge Theories}\label{section:3}

\subsection{A simple example of non-Abelian clouds} 
\label{sc:simplest}

\begin{figure}[thb]
\begin{center}
\includegraphics[width=6cm,clip]{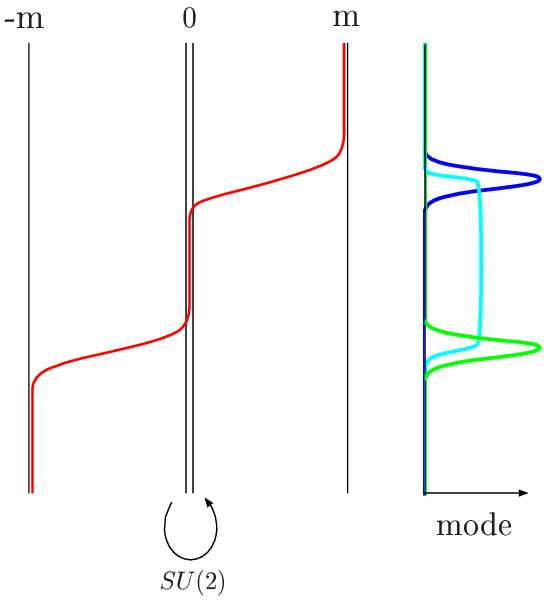}
\end{center}
\caption{\small
D-brane picture for a domain wall with non-Abelian clouds.
}
\label{fig5a}
\end{figure}

\begin{figure}[htb]
\begin{center}
\includegraphics[height=5.5cm]{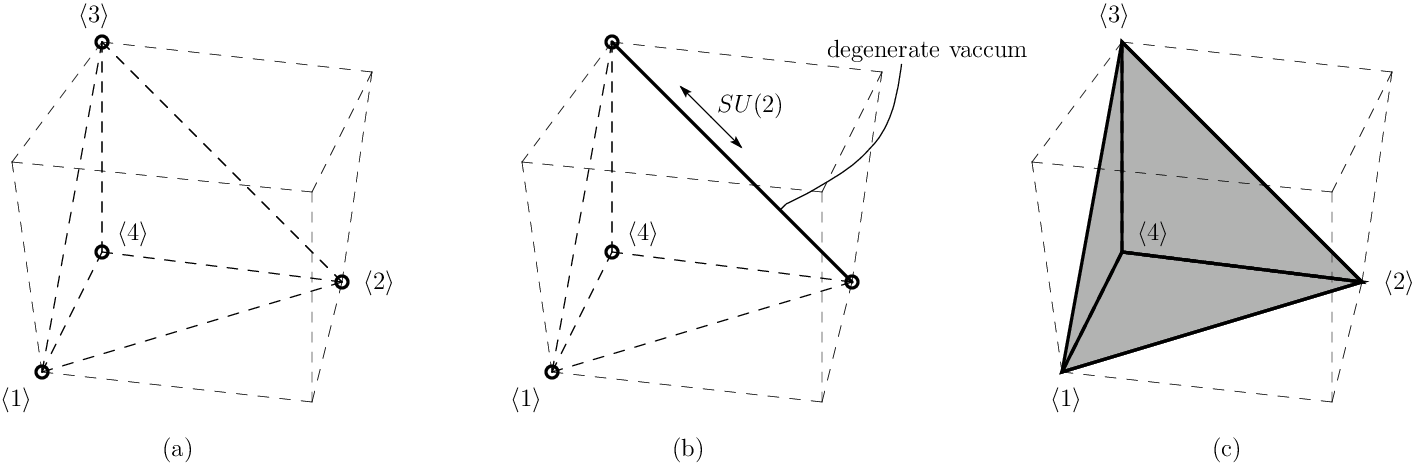}
\end{center}
\caption{\small
Vacua for various cases of mass configurations 
plotted in the three-dimensional space of Higgs 
fields $h_i^2, i=1,2,3$ with $\sum_{i=1}^{4}h_i^2=1$. 
(a) non-degenerate massive vacua 
(b) massive degenerate and non-degenerate vacua 
(c) massless vacuum
}
\label{cp4}
\end{figure}

Let us see the non-Abelian clouds in a simple example of 
the Abelian gauge theory coupled with the $N_{\rm F}=4$ 
Higgs fields. 
The corresponding brane configuration is shown in Fig.~\ref{fig5a}.
The massless vacuum manifold is $T^\star{\bf C}P^3$ 
where the base manifold is parametrized by
\beq
{\bf C}P^3 = \left\{HH^\dagger = c\right\}/U(1),\quad
H = \sqrt c \left(h_1,\ h_2,\ h_3,\ h_4 \right),
\eeq
where the quotient is the overall $U(1)$. 
The vacuum manifold is expressed as 
(the inside and the surface of)  
a triangular pyramid 
in the 3 dimensional space 
$(|h_1|^2, |h_2|^2, |h_3|^3)$, 
as shown in Fig.~\ref{cp4} (a). 
When the mass matrix 
containing a 
small 
parameter $\epsilon$ 
($0\le \epsilon \in {\bf R}$) 
\begin{eqnarray}
M = {\rm diag}
\left(
m,\ \frac{m\epsilon}{2},\  - \frac{m\epsilon}{2},\ -m
\right)
\label{eq:mass}
\end{eqnarray}
is turned on, the vacuum manifold is lifted except for 
four points and the flavor symmetry breaks from $SU(4)$ 
to $U(1)^3$.
These discrete vacua are the four vertices of the pyramid 
shown in Fig.~\ref{cp4} (b). 
We label those vacua as $\left<A\right>$ 
$(A=1,2,3,4)$. 
The vacuum expectation value (VEV) of the vacuum 
$\left<A\right>$ is $h_B = \delta_{AB}$.
Taking a limit of $\epsilon \to 0$, the second and the third 
Higgs fields become degenerate so that the flavor 
symmetry enhances 
from $U(1)^3$ to 
$U(1)^2 \times SU(2) \in SU(4)$. 
There are two isolated vacua 
and one degenerate vacuum 
${\bf C}P^1 \simeq SU(2)/U(1)$ 
represented by a line connecting $\left<2\right>$ and 
$\left<3\right>$ as shown by a thick line in 
Fig.\ref{cp4}(c).  
We denote this degenerate vacuum as $\left<\hbox{2-3}\right>$.

There exist domain wall solutions interpolating 
vacua in the model with fully or partially 
non-degenerate Higgs masses. 
In the case of $N_{\rm C}=1$, the moduli matrix and 
the $V$-equivalence (\ref{eq:Vtransformation}) 
take the form of 
\beq
H_0 = \left( \phi_1,\ \phi_2,\ \phi_3,\ \phi_4 \right)
\sim \lambda \left( \phi_1,\ \phi_2,\ \phi_3,\ \phi_4 \right),
\quad \lambda \in {\bf C}^*. \label{eq:N1Vequivalence}
\eeq
In terms of the moduli matrix the vacua $\left<A\right>$ 
is described by
$\phi_B = \delta_{BA}$ for $B=1,2,3,4$.
Since we want to consider the domain wall interpolating
the vacua $\left<1\right>$ and $\left<4\right>$ 
(passing by $\left<2\right>,\left<3\right>$ on the way),
the parameter $\phi_1$ and $\phi_4$
should not be zero while $\phi_2,\phi_3$ can become zero. 
So the moduli space corresponding
to the multiple domain walls which connect 
$\left<1\right>$ and $\left<4\right>$ is
\beq
 {\cal M} \simeq 
 \left( {\bf C}^2 \times ({\bf C}^*)^2\right) // {\bf C}^* 
 \simeq  {\bf C}^* \times {\bf C}^2, 
  \label{eq:moduli-simplest}
\eeq
where double slash denotes identification by 
the $V$-transformation.
Here the part ${\bf C}^* \simeq {\bf R} \times U(1)$ 
represents the translational modulus 
and the associated phase modulus.

\begin{figure}[htb]
\begin{center}
\begin{tabular}{ccc}
\includegraphics[height=5.5cm]{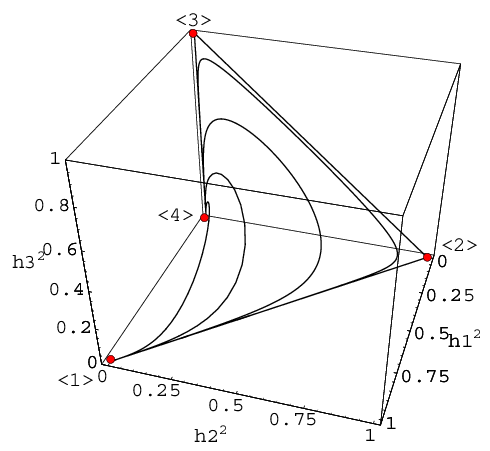} & \quad &
\includegraphics[height=5.5cm]{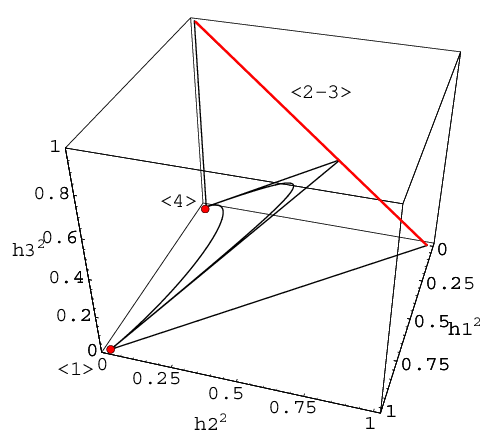} \\
(a) non-degenerate mass & \quad & (b) degenerate mass
\end{tabular}
\end{center}
\caption{\small
Domain wall trajectories in the target space ${\bf C}P^3$ 
for non-degenerate mass (a) and for degenerate mass (b). 
}
\label{cp3}
\end{figure}

When we take the gauge coupling $g$ to infinity, 
the model reduces to a nonlinear sigma model 
whose target space is the Higgs branch of 
the vacua in the original theory. 
To make the discussion simple, we take this limit for a while.
One benefit to consider the nonlinear sigma model is 
that the BPS equations are analytically solved. 
In fact the solutions are expressed as 
\cite{Isozumi:2004va} 
\begin{eqnarray}
H = \frac{1}{\sqrt \Omega_0} H_0e^{My}\quad
{\rm with}\quad
\Omega_0 \equiv H_0 e^{2My} H_0^\dagger. 
\end{eqnarray}
A domain wall solution corresponds to a trajectory 
connecting the vertex $\left<1\right>$ and 
$\left<4\right>$. Flows from $\left<1\right>$ to 
$\left<4\right>$ inside the pyramid
 are shown in Fig.~\ref{cp3}.

\begin{figure}[htb]
\begin{center}
\begin{tabular}{ccccc}
\includegraphics[height=2.5cm]{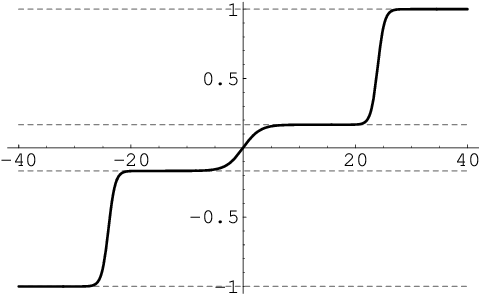} & \qquad \quad &
\includegraphics[height=2.5cm]{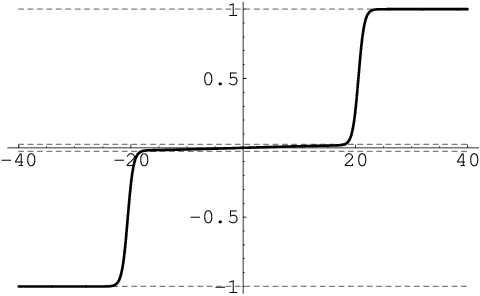} & \qquad \quad &
\includegraphics[height=2.5cm]{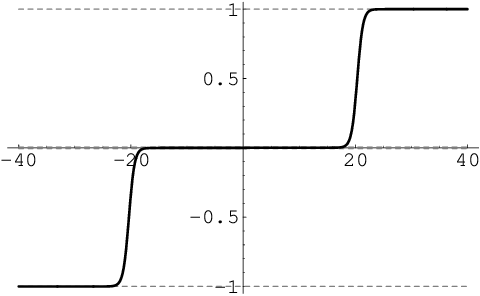} \\
\includegraphics[height=2.5cm]{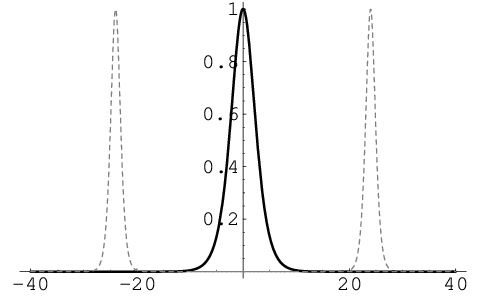} & &
\includegraphics[height=2.5cm]{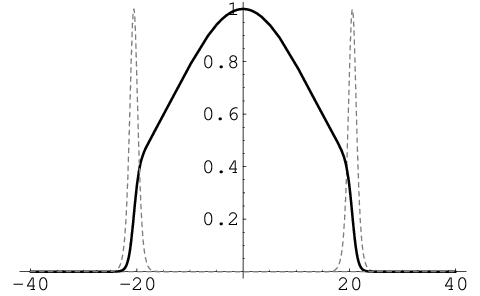} & &
\includegraphics[height=2.5cm]{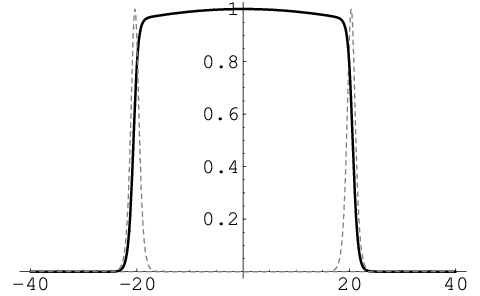}\\
(a) $\epsilon=1/3$ & & (b) $\epsilon=1/20$ & & (c) $\epsilon=1/100$
\end{tabular}
\end{center}
\caption{\small
Configuration of $\Sigma$ 
(first row) 
and density of the K\"ahler metric 
of $\varphi_1$, $\varphi_2$
and $\varphi_3$ (second row). 
Moduli parameters are
$(\varphi_1,\varphi_2,\varphi_3)=(20,0,-20)$ and $m=1$.
}
\label{fig8}
\end{figure}

Physical meaning of the moduli parameters becomes much 
clearer by 
using the $V$-equivalence relation (\ref{eq:N1Vequivalence}) 
to fix the form of the moduli matrix as 
\begin{eqnarray}
H_0 =  \left( 1,\ e^{\varphi_1},\ 
e^{\varphi_1+\varphi_2},\ e^{\varphi_1+\varphi_2+\varphi_3} \right).
\label{eq:mm_3wall}
\end{eqnarray}
Furthermore, one may be visually able to see the ``kink" 
configuration in the profile of 
the field $\Sigma=(1/2)\partial_y \log \Omega_0$.
In the vacuum region $\left<A\right>$ the function $\Sigma(y)$ takes the value
$\Sigma = m_A$. Several solutions are shown in Fig.~\ref{fig8}.
The domain wall positions can be roughly read from the 
moduli matrix in Eq.~(\ref{eq:mm_3wall}) as 
\begin{eqnarray}
\frac{y_+}{L_+} = \varphi_1 + \varphi_1^*,\quad
\frac{y_0}{L_0} = \varphi_2 + \varphi_2^*,\quad
\frac{y_-}{L_-} = \varphi_3 + \varphi_3^*, \label{eq:3positions}
\end{eqnarray}
where $y_+$ is the position of the right wall and $y_0,y_-$ 
are the positions for the middle and the left walls, 
respectively. 
Here $L_{\pm,0}$ stands for the width of each wall 
\begin{eqnarray}
L_+ \equiv \frac{2}{m(2-\epsilon)}, \quad 
L_0 \equiv \frac{1}{m\epsilon},\quad 
L_- \equiv \frac{2}{m(2-\epsilon)}. 
\end{eqnarray}
This rough estimation is, of course, valid only for well 
separated walls whose positions are aligned as 
$y_- \ll y_0 \ll y_+$, see Fig.~\ref{fig8} (a). 
Each domain wall is accompanied by a complex moduli 
parameter $\varphi_i$ whose real part is related to the 
wall position and 
imaginary part 
is the $U(1)$ internal symmetry 
(the Nambu-Goldstone mode associated with the broken $U(1)$ 
flavor symmetry).  

To argue symmetry aspects of the moduli parameters, 
first let us consider a model which has completely 
non-degenerate masses and domain walls interpolating 
between those vacua. The 
global symmetry explicitly breaks
from $SU(4)$ to $U(1)^3 \subset SU(4)$. 
We take, as the unbroken 
global symmetries, $U_1(1),U_2(1)$ and $U_3(1)$
with generators 
${\rm diag}\left( 1, -1, -1, 1\right)$,
${\rm diag}\left( 1, 0, 0, -1\right)$,
and
${\rm diag}\left( 0, 1,-1, 0 \right)$
respectively.
Each vacua $\left<A\right>$ preserves all of these symmetries.
However, once domain walls connecting those vacua appear, 
they break all or a part of these symmetries. For example, the moduli matrix $H_0 = (1,0,0,\phi_4)$ corresponding to a domain wall connecting two vacua 
$\left<1\right>$ and $\left<4\right>$ breaks $U_2(1)$ but still preserves 
$U_1(1)$ and $U_3(1)$. 
Here note that overall phase can be 
absorbed by the $V$-transformation (\ref{eq:N1Vequivalence}). 
Therefore the phase of the moduli parameter $\phi_4$ corresponds to nothing but the broken global symmetry $U_2(1)$.
This implies the Nambu-Goldstone mode localizes 
around the domain wall as we saw above.
For the moduli matrix $H_0=(1,\phi_2,0,\phi_4)$, 
which corresponds to two domain walls connecting three vacua 
$\left<1\right> \to \left<2\right> \to \left<4\right>$, 
the symmetry $U_3(1)$ in addition to $U_2(1)$ breaks while a combination of 
$U_1(1)$ and $U_3(1)$ is still preserved.
Moreover, when we turn on the third element in the moduli matrix as $H_0=(1,\phi_2,\phi_3,\phi_4)$,
the third vacuum region appears and then the configuration has three domain walls connecting four vacua 
$\left<1\right> \to \left<2\right> \to \left<3\right> \to \left<4\right>$.
In this case all of $U(1)^3$ are broken by the domain walls, so that 
corresponding three Nambu-Goldstone modes appear.
These three Nambu-Goldstone modes are described by
imaginary parts of $\log \phi$, 
which are combined with the three positions (\ref{eq:3positions}), 
to form three complex coordinates of the moduli space 
${\bf C}^2 \times {\bf C}^*$.

Next we consider a limit where the 
second and the third 
masses are degenerate 
($\epsilon\to0$ in the mass matrix (\ref{eq:mass})).
In the limit the global symmetry 
$U_1(1)\times U_2(1)\times U_3(1)$ is enhanced to 
$U_1(1)\times U_2(1)\times SU(2)$. 
At the same time, 
the degenerate vacuum $\left<\hbox{2-3}\right>$ appear 
instead of the two isolated vacua $\left<2\right>$ and
$\left<3\right>$ 
as shown in Fig.~\ref{cp4} (c). 
At the degenerate vacuum, $U_1(1),U_2(1)$ are preserved 
but $SU(2)$ is broken to $U_3(1)$. 
Therefore the degenerate vacuum $\left<2\text{-}3\right>$ is 
$SU(2)/U_3(1) = {\bf C}P^1$. 
Non-vanishing $\phi_4\not=0$ causes 
the wall interpolating two vacua $\left<1\right> \to \left<4\right>$ 
and breaks only $U_2(1)$ again. 
Once the degenerate vacuum appears in the configuration such as 
two domain walls connecting vacua like 
$\left<1\right> \to \left<2\text{-}3\right> \to \left<4\right>$,
the breaking pattern of the global symmetry becomes different 
from that in the case of fully non-degenerate masses. 
The moduli matrix $H_0 = (1,\phi_2,\phi_3,\phi_4)$ describes such domain walls. Note that the second and the third
elements breaks $SU(2)$ completely. The global symmetry $U_1(1) \times U_2(1) \times SU(2)$ are broken to $U(1)$ which 
is a mixture of $U_1(1)$ and $H \in SU(2)$. 
Emergence of the second wall and further $U(1)$-symmetry breaking 
are related to the facts that $|\phi_2|^2+|\phi_3|^2\not=0$ 
and $\phi_4\not=0$.
These facts imply that the modes corresponding to the 
two broken $U(1)$'s localize around the walls accompanied by the 
two position moduli and the mode corresponding to 
$SU(2)/H$ have support in a region around 
the degenerate vacuum $\left<2\text{-}3\right>$.
This is consistent with the observation from the view point of the 
D-brane picture Fig.~\ref{fig5a}. 
We can count the number of the moduli parameters as follows.
Two real parameters $\{|\phi_2|^2+|\phi_3|^2,|\phi_4|^2\}$ 
correspond to the positions of the two walls 
whereas remaining four parameters 
correspond to the broken global symmetry 
$U_1(1) \times U_2(1) \times SU(2)/U(1)$. 
This is again consistent with
${\rm dim}_{{\bf R}} \left({\bf C}^2 \times {\bf C}^*\right)$.

In the Fig.~\ref{fig8} we showed domain wall configurations 
of the three domain walls connecting the four vacua. 
As the parameter $\epsilon$ 
decreases, 
the width of 
the middle domain wall connecting the vacua 
$\left<2\right>$ and $\left<3\right>$ becomes 
broad and the tension of the wall becomes small 
since they are proportional to $1/\epsilon$ 
and $\epsilon$ respectively. 
When the width of the middle wall becomes lager than the 
separation of two outside walls, 
$L_0 \gtrsim y_+ - y_-$, 
we can no longer see the middle wall. 
The density of the K\"ahler metric for the moduli 
parameters $\varphi_1$, $\varphi_2$ and $\varphi_3$ in the strong 
gauge coupling limit are shown in the second row of 
Fig.~\ref{fig8}. 
The K\"ahler potential in the strong coupling limit 
is given by 
$K = c\int dy\ \log \Omega_0$ \cite{Eto:2006uw}. 
When three walls are well isolated as Fig.~\ref{fig8} (a), 
three modes corresponding to the moduli parameters 
$\varphi_1$, $\varphi_2$ and $\varphi_3$ 
are localized on the 
respective domain walls.  
As $\epsilon$ decreases, the density of the K\"ahler metric of 
$\varphi_2$ is no longer localized but is stretched 
between two outside domain walls. 
In the limit where $\epsilon\to 0$ the physical meaning of 
$\varphi_2$ as the position and the internal phase 
associated with the middle domain wall should be 
completely discarded. 
Instead, $\varphi_2$ gives the non-Abelian cloud which comes 
from the flat direction ${\bf C}P^1$ of the degenerate 
vacua $\left<2\text{-}3\right>$. 
For each fixed moduli parameters 
$\varphi_1, \varphi_2, \varphi_3$, 
the domain wall solution as a function of $y$ sweeps out 
a trajectory in the target space ${\bf C}P^3$. 
These domain wall trajectories are shown for various 
values of moduli parameters in Fig.~\ref{cp3}: 
non-degenerate mass case (a) and degenerate mass case (b). 
For degenerate mass case, the trajectories 
do not go out from the triangular plane whose 
vertices are $\left<1\right>, \left<4\right>$ and one 
point on the edge between $\left<2\right>$ and 
$\left<3\right>$.

\subsection{Effective action of 
non-Abelian clouds and their dynamics
}\label{subsec:efac}

Next we construct the effective action for the non-Abelian clouds 
with leaving the gauge coupling to be finite. 
In this subsection 
we consider a more general model with $N_{\rm F}$ flavors with masses
\beq
 M = (m_1,0,0,\cdots,0,-m_2)  ,\quad m_1,m_2>0.
\eeq 
There exist two isolated points of vacua and one
continuously degenerate vacua ${\bf C}P^{N_{\rm F}-3}$.

This model admits 
two domain walls interpolating between two isolated vacua at 
$y=-\infty$ to $y=+\infty$ with the degenerate
vacua ${\bf C}P^{N_{\rm F}-3}$ 
between the two domain walls. 
The full moduli space is 
\beq
 {\cal M} \simeq {\bf C}^* \times {\bf C}^{N_{\rm F}-2}.
 \label{eq:abelian_model_moduli}
\eeq
In the following we do not consider the ${\bf C}^*$ 
corresponding to the center of the mass 
and the overall phase. 
Then let us take the moduli matrix 
\beq
 H_0 = (1,\phi_2,\phi_3,\cdots,\phi_{N_{\rm F}-1},1).
\eeq 
The positions of the two walls 
can be estimated as 
\beq
y_1 = \frac{1}{2m_1} \log |\vec\phi|^2,\quad
y_2 = -\frac{1}{2m_2} \log |\vec\phi|^2,
\eeq
with a vector $\vec \phi \equiv (\phi_2,\phi_3,\cdots, \phi_{N_{\rm F}-1})$.
Notice that we have fixed the center of mass of the two walls
as $m_1y_1 + m_2y_2 = 0$.
The distance of the two walls is defined as
\beq
R = y_1 - y_2 = \frac{1}{\mu} \log|\vec\phi|^2,\quad
\mu \equiv \frac{2m_1m_2}{m_1+m_2}.
\eeq
The function $\Omega_0(y)$ in the master equation (\ref{eq:master}) 
in this case is given by
\begin{eqnarray}
 \Omega_0  = c^{-1}\left(e^{2m_1y} + |\vec\phi|^2 + e^{-2m_2y}\right).
\end{eqnarray}
Although we have to solve the master equation 
(\ref{eq:master}) to obtain the explicit expression 
of the quantity $\Omega$, we do not need it for the later analysis:  
it is easy to see
that the K\"ahler potential (\ref{eq:formal-Kahler}) 
depends only 
on\footnote{
The K\"ahler potential of this type was studied in 
\cite{Higashijima:2001vk} 
where the Ricci-flat metric on a line bundle over the projective space 
was obtained by enforcing the Ricci-flat condition. 
Here, the metric does not have to be Ricci-flat of course.
} 
$\mu R = \log|\vec\phi|^2$ 
\begin{eqnarray}
 K(\phi,\phi^*) = f\left( \mu R \right).
 \label{eq:f}
\end{eqnarray}
We give asymptotic form of the function $f$, below.
The effective action is obtained from the K\"ahler 
potential via the K\"ahler metric as 
$
L_{\rm eff} = 
K_{ij^*} \p_\mu \phi^i \p^\mu \phi^{j*}
$.
After changing the variables as
\beq
\vec \phi = e^{(\mu R + i\xi)/2} \vec n, \hs{10}
|\vec n|^2 = 1,
\eeq
we can obtain the following expression
\beq
{\cal L}_{\rm eff} = 
\frac{1}{4} f''(\mu R) \left[
\mu^2 (\p_\mu R)^2 + (\p_\mu \xi - 2i \vec n^\dagger\p_\mu \vec n)^2
\right] + 
f'(\mu R) \left[ |\p_\mu \vec n|^2 - |\vec n^\dagger \p_\mu \vec n|^2\right].
 \label{eq:eff_ac_abelian_model}
\eeq
Here the complex vector $\vec n$ consists of the 
coordinates of the vacua ${\mathbf C}P^{N_{\rm F}-3}$ 
between the two walls, that is,  
the non-Abelian clouds.\footnote{ 
The target space metric of the effective Lagrangian 
(\ref{eq:eff_ac_abelian_model})
locally looks like a complex line bundle over 
${\mathbf C}P^{N_{\rm F}-3}$, namely 
${\cal O}(-1) \rightarrow {\mathbf C}P^{N_{\rm F}-3}$.
However it does not hold for $R \to - \infty$ (coincident walls) 
where the metric tends to a single point 
as found in Eq.~(\ref{eq:asympt_f}), below.
Therefore the base space ${\mathbf C}P^{N_{\rm F}-3}$ 
of the bundle 
is blown down to a point to obtain 
${\bf C}^{N_{\rm F}-2}$ 
in the moduli space (\ref{eq:abelian_model_moduli}).
}

\begin{figure}[thb]
\begin{center}
\begin{tabular}{cc}
\includegraphics[width=8cm]{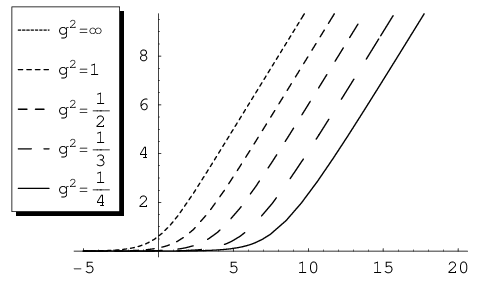} &
\includegraphics[width=8cm]{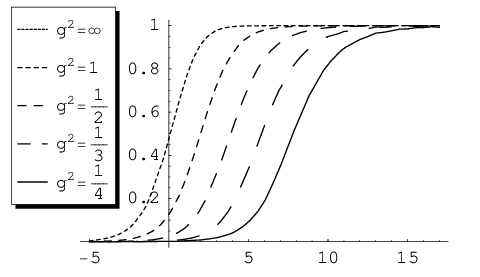} \\
(i)~\,$f'(\mu R)$ & (ii)~\,$f''(\mu R)$ 
\end{tabular}
\end{center}
\caption{\small Typical profiles of the functions $f'(\mu R)$ and $f''(\mu R)$ with $c=1,~m_1=m_2=1$. The function $f(\mu R)$ is numerically calculated for the gauge coupling $g^2=\infty,\,1,\,\frac{1}{2},\,\frac{1}{3},\,\frac{1}{4}$.}
\label{f(mu R)}
\end{figure}

Since two walls become independent as they are separated by a large distance,
the kinetic term of the relative distance $R$ should be a free action 
${\cal L}_{\rm free} = \frac{\mu c}{4}(\p_\mu R)^2$ 
for sufficiently large $R$. Note that the coefficient $\mu c/4$ is
calculated by using Eq.(\ref{eq:inertial-mass}). 
Furthermore, the K\"ahler metric written in the moduli fields $\phi^i$ which
are original entries in the moduli matrix should be smooth everywhere, especially at $|\vec \phi| = 0$ ($R\to-\infty$). From these two facts we can find the asymptotic behavior of the function $f(\mu R)$ as
\beq
f(\mu R) ~=~ \left\{ 
\begin{array}{ccl} 
\displaystyle \frac{c \mu R^2}{2} - (d_1 + d_2) c \mu R + \mathcal O(1) & & \text{for}\quad R \to \infty \\ 
\displaystyle \phantom{\frac12} A \, e^{\mu R} + \mathcal O(e^{2 \mu R}) & \hs{10} & \text{for} \quad R \to -\infty
\end{array} 
\right., 
 \label{eq:asympt_f}
\eeq
where $d_1 \equiv m_1/g^2 c$ and $d_2 \equiv m_2/g^2 c$ are half of the widths of walls and $A$ is a constant determined by solving the BPS equations. 
The derivation of the subleading term for well-separated walls, which is proportional to $R$, is given in Appendix \ref{appendix:1}.
Note that in the region of $R<d_1+d_2$, the parameter $R$ no 
longer has the meaning of the distance between the walls  
and the two walls are nearly compressed into one wall. 
Especially at $|\vec \phi| =0~(R \rightarrow - \infty)$,
the two walls are completely compressed and
the degrees of freedom of the non-Abelian clouds
between the two walls disappear
with the shrinking of ${\mathbf C}P^{N_{\rm F}-3}$ to a point. 
Fig.\,\ref{f(mu R)}-(i) and-(ii) show the typical profiles 
of the functions $f'(\mu R)$ and $f''(\mu R)$ for various 
values of the gauge coupling constant.

In order to consider the dynamics in detail, 
let us concentrate on the minimal case of $N_{\rm F} = 4$ 
and consider 
the kinks in the $d=1+1$ gauge theory with 
the $(1+0)$ dimensional world-volume 
( 
the world-volume is time only $\mu=0$
) in the rest of this section. 
It is convenient to redefine the parameters as
\beq
(\phi_2,\ \phi_3) = \vec \phi 
= e^{(\mu R + i\xi)/2} \vec n 
= e^{(\mu R + i\xi)/2} \left(e^{i\varphi/2} 
\cos \frac{\theta}{2},\ e^{-i\varphi/2} \sin \frac{\theta}{2}\right).
\eeq
The Lagrangian in these coordinates takes the form 
\beq
L_{\rm eff} &=& 
\frac{f''(\mu R)}{4} \left[ \mu^2 \dot R^2 
+ \left( \dot \xi + \cos \theta \, \dot \varphi \right)^2\right] 
+ \frac{f'(\mu R)}{4} 
\left( \dot \theta^2 + \sin^2 \theta \, \dot \varphi^2 \right) \\
&\rightarrow& \frac{c}{4\mu} \left[ \mu^2 \dot R^2 
+ \left( \dot \xi + \cos \theta \, \dot \varphi \right)^2\right] 
+ \frac{c}{4} \left( R - d_1 - d_2 \right) 
\left( \dot \theta^2 + \sin^2 \theta \, \dot \varphi^2 \right) .
\label{eq:eff_N=2}
\eeq 
The asymptotic behavior of the coefficients 
of kinetic terms of $\theta$ and $\varphi$ reflects the 
fact that the wave functions of the non-Abelian clouds are 
extending in the interval between the walls which is 
effectively reduced by the widths of the walls. 
As mentioned above, the K\"ahler potential depends only on 
$\mu R = \log |\vec \phi|^2$, so that there exist four 
conserved quantities defined by 
\beq
Q &=& \frac{i}{2} K_{i j^\ast} 
\left( \dot{\phi}^{j\ast} \phi^i 
- \dot \phi^i \phi^{j\ast} \right), \\
q_a &=& \frac{i}{2} K_{i j^\ast} \left( \dot{\phi}^{j\ast} 
{\left( \sigma_a \right)^i}_k \phi^k 
- \dot \phi^i \phi^{k\ast} {\left( \sigma_a \right)_k}^j \right),
\eeq
where $\sigma_a~(a=1,2,3)$ are the Pauli matrices. 
These conserved charges originate from $U(2)$ symmetry 
which rotates the complex vector $\vec \phi$. 
Note that they are not independent, but related as 
\beq
Q = \frac{1}{|\vec{\phi} |^2} 
\left( \phi^\ast \sigma_a \phi \right) q_a.
\eeq
By using these conserved charges, 
we can rewrite the Lagrangian as 
\beq
\tilde L_{\rm eff} ~=~ \frac{\mu^2}{4} f''(\mu R) 
\dot R^2 - \frac{Q^2}{f''(\mu R)} 
- \frac{q_a q_a - Q^2}{f'(\mu R)} .
\eeq

Let us consider the dynamics of kinks when  
these conserved charges take non-zero values.
Then 
we have two types of potential between two walls: one is 
given by $V_1(R) = 4Q^2/f''(\mu R)$, which exponentially 
approaches to a constant as the distance $R$ become larger, 
and the other is $V_2(R) = 2(q_a q_a - Q^2)/f'(\mu R)$, 
which behaves as $V_2(R) \approx 1/R$ for large $R$. 
The former type of the potential also exists in the case 
of fully non-degenerate masses. 
The novel feature here is the existence of the potential 
$V_2(R)$ which leads to a long range repulsive force 
between two walls. 
Physical interpretation of these potentials is quite 
interesting;
In the case of fully non-degenerate masses, there are 
only massive modes which propagate between two walls, 
so that the potential falls off rapidly for large 
$R \gg 1/\mu$. 
In the case of degenerate masses, 
we have some massless Nambu-Goldstone modes propagating 
between two walls. 
They are nothing but the non-Abelian clouds 
and mediate the long range repulsive force: 
the motion in the internal space induces a 
repulsive force between the two kinks.

\subsection{Kink bound state 
stabilized by non-Abelian clouds}
\label{subsec:kink-bound}
We can let the degenerate mass to split by giving imaginary 
masses for the Higgs scalar fields. 
Then an attractive force between the two kinks is induced, 
and  
a bound state of the two kinks can be formed. 
When we add the additional masses of 
the scalar fields in the original theory as 
\beq
M \to M + i\tilde M = {\rm diag}\left(m_1, i\tilde m/2,-i \tilde m/2, -m_2 \right),\quad \tilde m > 0,\label{eq:additional-masses}
\eeq 
the vacuum manifold is lifted and the 
continuous degeneracy of the vacua disappears. 
These imaginary masses make domain walls Abelian.
We can, however, easily keep a part of continuous
degeneracy of vacua, by extending the system to a model
with real and imaginary masses in the case of $N_{\rm F}>4$.
In that case, domain walls remain non-Abelian. 
Here we consider (\ref{eq:additional-masses}) for simplicity. 
For a small $\tilde m \ll \mu$, 
the additional masses induce a potential which is given by 
the squared norm of the Killing vector 
$k = \tilde m \p_\varphi$ on the moduli space as 
\beq
V_{\rm eff} =
\frac{\tilde m^2}{4} \left[ f''(\mu R) \cos^2 \theta 
+ f'(\mu R) \sin^2 \theta \right]. 
\eeq
This is an attractive potential with minimum at 
$|\vec \phi|^2 = e^{\mu R} =0$, namely two walls tend to 
be compressed into one wall. 
Once the additional masses are turned on, 
not all 
 charges $Q$ and $q_a$ are conserved, but $Q$ and $q_3$ 
are left to be conserved. 
The charges $Q$ and $q_3$ are conjugate momenta of $\xi$ 
and $\varphi$, respectively 
\beq
\frac{\p L_{\rm eff}}{\p \dot \xi} &=& Q ~=~ 
\frac{f''(\mu R)}{2} (\dot \xi + \cos \theta \, \dot \varphi), \\
\frac{\p L_{\rm eff}}{\p \dot \varphi} 
&=& q_3 ~=~ \frac{f''(\mu R)}{2} \cos \theta \, 
(\dot \xi + \cos \theta \, \dot \varphi) 
+ \frac{f'(\mu R)}{2} \sin^2 \theta \, \dot \varphi .
\eeq
Therefore, we effectively obtain the following potential
\begin{eqnarray}
\tilde V_{\rm eff} &=& \frac{\tilde m^2}{4} \left[ f''(\mu R) \cos^2 \theta + f'(\mu R) \sin^2 \theta \right] + \frac{Q^2}{f''(\mu R)} + \frac{(q_3-\cos \theta \, Q)^2}{f'(\mu R) \sin^2 \theta}  .
\end{eqnarray} 
The potential is composed of four terms with two different 
types of asymptotic behaviors, namely long-range and 
short-range forces: the first and third terms exponentially 
approach to constants for large $R$, while the second and 
fourth terms are proportional to $R$ and $1/R$ respectively. 

The effective potential is bounded from below 
\begin{eqnarray}
\tilde V_{\rm eff} 
&\geq& \tilde m |q_3|.
\end{eqnarray} 
This lower bound of the effective potential is saturated 
if $R$ and $\theta$ satisfy 
\beq
\frac{\tilde m}{2} f''(\mu R) \cos \theta 
= \eta Q, \hs{10} \frac{\tilde m}{2} f'(\mu R) \sin^2 \theta 
= \eta (q_3-\cos \theta \, Q), \hs{5} 
\eta \equiv {\rm sign} (q_3).
\eeq
The solution of these equations shows various properties 
for given values of the conserved charges $Q$ and $q_3$. 
In the following we consider two cases, 
1) $|Q|=|q_3|$ and 2) $Q=0$.

1) As an example, let us consider the case where the 
the absolute values 
of two charges are the same $|Q|=|q_3|$. 
In this case, the relative distance $R$ and the phase 
$\theta$ are stabilized at 
\begin{eqnarray}
\theta = \left\{
\begin{array}{ccl} 
0 & & \text{for}~~ Q ~ = ~~ q_3 \\
\pi & & \text{for}~~ Q ~ = -q_3 
\end{array}\right.,~~
\ R = R_0 \hs{5} \text{with}~~ f''(\mu R_0) = \frac{2|q_3|}{\tilde m}.
\label{eq:Q1}
\end{eqnarray} 
In this case, the positions of two walls are stabilized at 
the points where the two short-range forces balance. 
Because of this short range force the two walls stabilize 
with either small separation $R \lesssim 1/\mu$ or large 
separation $R \gtrsim 1/\mu$ with exponentially weak 
binding force. 
The squared mass of the fluctuation of the relative 
distance is given by 
$m^2_{\delta R} = (\tilde m f^{(3)}(\mu R_0)/f''(\mu R_0))^2$, 
which becomes exponentially small for large $R_0$. 
Especially, if two wall system have too much conserved 
charge $|q_3| \geq {\rm max} (f''(\mu R)) = \tilde m c/\mu$, 
an instability appears: the minimum of the potential 
disappears to infinity $R \rightarrow \infty$ (runaway 
potential). 
This type of the stabilized wall also exists as $Q$-walls 
(dyonic walls) in models with fully non-degenerate masses 
\cite{Lee:2005sv,Eto:2005sw}. 
Actually, the corresponding configuration to the solution 
(\ref{eq:Q1}) can be obtained by embedding the $Q$-wall 
solution in a model with non-degenerate masses into the 
model we are now considering.
\\
2) 
Another example is the case with $Q=0$. 
In this case, the relative distance $R$ and the 
phase $\theta$ are stabilized 
at
\beq
\theta =  
\frac{\pi}{2}, \hs{5} R = \tilde R_0, \hs{5} \text{with}~~ f'(\mu \tilde R_0) = \frac{2 |q_3|}{\tilde m}.
\eeq
The 
two walls are stabilized at 
the point where 
the two long-range forces balance. 
Because of these long-range forces the positions of the 
two walls can be stabilized with a large separation. 
The squared masses 
of the fluctuations around the minimum of the potential 
are given by 
$m_{\delta R}^2 = m_{\delta \theta}^2 
= \tilde m^2 f''(\mu \tilde R_0) / f'(\mu \tilde R_0)$, 
which behave as $1/\tilde R_0$ for large relative distance. 
There is no instability even if $|q_3| \gg \tilde m c/\mu$, 
since $f'(\mu R)$ grows linearly for large $R$. 
These properties are in contrast to the case of fully 
non-degenerate masses. 
All these differences between fully non-degenerate and 
degenerate masses originate from the existence of 
the non-Abelian clouds in the degenerate case 
which give the long-range interactions. 

Finally let us make a comment on supersymmetry.
The masses 
(\ref{eq:additional-masses}) for Higgs fields 
(hypermultiplets) are possible in dimensions 
$3+1$ or less. 
The stable configurations of the Q-walls (dyonic walls) 
considered in this subsection are 1/4 BPS states 
\cite{Lee:2005sv,Eto:2005sw}.

\section{The Generalized Shifman-Yung Model}
\label{sc:GSY}

\subsection{The model and its vacua}
\label{sc:strong-coupling}

\begin{figure}[thb]
\begin{center}
\begin{tabular}{ccc}
\includegraphics[width=5cm,clip]{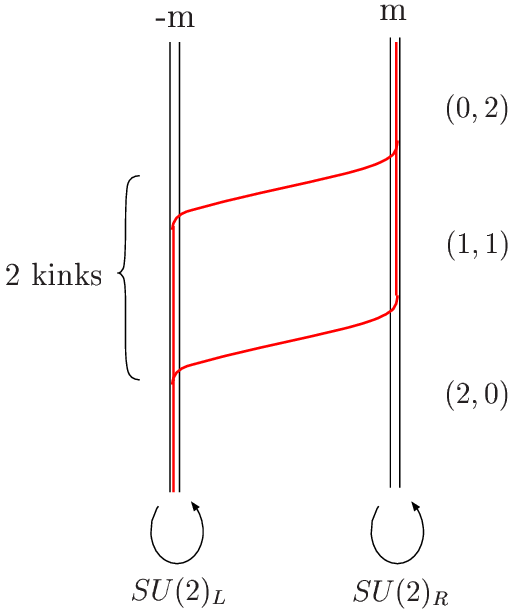} & &
\includegraphics[width=5cm,clip]{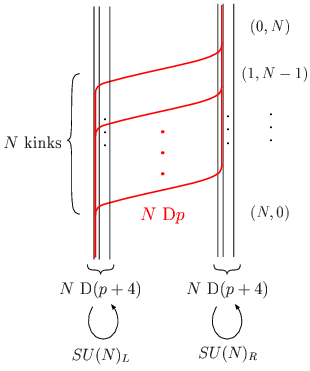} \\
(a) & \hs{15} & (b) 
\end{tabular}
\caption{\small
The kinky D-brane configurations for the SY model 
(a) 
and for the 
GSY model (b). 
\label{fig:D-brane-SY}
}
\end{center}
\end{figure}

In this section we consider non-Abelian gauge theory 
with degenerate masses of the Higgs fields. 
The simplest such situation may be provided by two sets of two 
degenerate mass parameters of the Higgs fields. 
Previously considered model is the $U(2)$ gauge theory 
with four Higgs fields in the fundamental representation 
with the mass matrix $M = {\rm diag}(m,m,-m,-m)$ 
\cite{Shifman:2003uh,Eto:2005cc}, 
which we call the Shifman-Yung Model. 
The model enjoys a flavor symmetry 
$SU(2)_{\rm L}\times SU(2)_{\rm R}\times U(1)_{\rm A}$. 
This model admits two domain walls  
which can pass through each other, 
in contrast to the Abelian gauge theory 
where walls do not pass through each other.
It has been demonstrated that 
the coincident domain wall 
configurations break the flavor symmetry to $SU(2)_{\rm V}$ 
and the Nambu-Goldstone bosons corresponding to
$[SU(2)_{\rm L}\times SU(2)_{\rm R}\times U(1)_{\rm A}]/ SU(2)_{\rm V} 
\simeq U(2)$ 
appear in the effective action on the walls. 
The symmetry breaking is the same as 
that of the chiral symmetry in hadron physics. 
The kinky D-brane configuration for 
this wall configuration 
is shown in Fig.~\ref{fig:D-brane-SY}-(a). 
Up to two D$p$-branes are allowed to lie inside D($p+4$)-branes 
by the s-rule \cite{HananyWitten97}. 

To generalize the non-Abelian flavor symmetry 
$SU(2)_{\rm L} \times SU(2)_{\rm R}$ 
to $SU(N)_{\rm L} \times SU(N)_{\rm R}$, 
we consider the $U(N)$ gauge theory 
with $N_{\rm F}=2N$ Higgs fields in the fundamental representation 
whose mass matrix is given by  
\begin{eqnarray}
M ~=~ m {\sigma _3 \over 2} \otimes {\bf 1}_N ~=~ {1\over 2}{\rm diag}(\overbrace{m,\cdots, m}^{N},\overbrace{-m,\cdots,-m}^{N}).
\end{eqnarray}
This system has a non-Abelian flavor symmetry 
$SU(N)_{\rm L}\times SU(N)_{\rm R}\times U(1)_{\rm A}$. 
Since we have only two mass parameters $m$ and $-m$, possible 
vacua are classified by an integer $0 \le k \le N$: 
in the $k$-th vacua, there is a configuration in which 
$k$ flavors of the first half 
and $N-k$ flavors of the latter half take non-vanishing 
values and then $\Sigma $ and $H$ are 
\begin{eqnarray}
 \Sigma \big|_{\rm vacuum}&=&{1\over 2} \, {\rm diag}
(\overbrace{m,\cdots, m}^{k},\overbrace{-m,\cdots,-m}^{N-k}),\nn \\
H\big|_{\rm vacuum}&=&\sqrt{c}\left(
\begin{array}{cccc}
 {\bf 1}_k& {\bf 0}       & {\bf 0}_k &{\bf 0} \\
 {\bf 0}  & {\bf 0}_{N-k} & {\bf 0} &{\bf 1}_{N-k}
\end{array}
\right).
\label{eq:kthvacuum}
\end{eqnarray}
This vacuum is labeled as $(k,N-k)$. 
The flavor symmetry $SU(N)_{\rm L}$ is broken down to 
$SU(k)_{\rm C+L}\times SU(N-k)_{\rm L}\times U(1)_{\rm C+L}$,
and $SU(N)_{\rm R}$ 
is broken down to 
$SU(k)_{\rm R}\times SU(N-k)_{\rm C+R}\times U(1)_{\rm C+R}$.
Therefore 
in this vacua there emerge 
$4k(N-k)$ Nambu-Goldstone modes, 
which parametrize 
the direct product of two Grassmann manifolds,
\begin{eqnarray}
G^{\rm L}_{N,k}\times G^{\rm R}_{N,k} . 
\label{eq:intra-vacua} 
\end{eqnarray}
Consequently 
the number of 
the discrete components of
the vacua is $N+1$ in this system. 

The unbroken symmetries of the vacua 
$(N,0)$ and $(0,N)$ which we consider in the next subsections 
as the boundary condition of domain walls, are 
$SU(N)_{\rm C+L} \times U(1)_{\rm C+L} \times SU(N)_{\rm R}$ 
and  
$SU(N)_{\rm L} \times SU(N)_{\rm C+R} \times U(1)_{\rm C+R}$, 
respectively.

\subsection{General solution of domain walls }\label{subsec:Gen-sol}
Walls are obtained by interpolating between a vacuum 
at $y=-\infty$ and another vacuum at $y=\infty$. 
The boundary conditions at both infinities define topological 
sectors. 
For a given topological sector, we may find several walls. 
The maximal number of walls in this system is $N$, 
which are obtained for the following maximal topological 
sector 
\begin{eqnarray}
 H=\left\{\begin{array}{cc}
\sqrt{c}({\bf 1}_N,{\bf 0}_N)& {\rm at~} y=+\infty \\
\sqrt{c}({\bf 0}_N,{\bf 1}_N)& {\rm at~} y=-\infty \\
	    \end{array}\right. . 
\label{eq:bound_cod}
\end{eqnarray}
In this case, the moduli matrix $H_0$
can be set into the following form without loss of generality:
\begin{eqnarray}
 H_0=\sqrt{c}({\bf 1}_N, e^\phi) \sim \sqrt{c}(e^{-\phi},\,{\bf 1}_N) ,
\end{eqnarray}
where $e^\phi $ is an element of $GL(N,{\bf C})$ and 
$\phi$ describes the moduli space of walls of this system, 
and the two forms are related by the $V$-transformation 
(\ref{eq:Vtransformation}). 
The $GL(N,{\bf C})$ matrix 
$e^\phi$ can always be rewritten as a product of 
a unitary matrix $U$ and 
a Hermitian matrix $e^{\hat x}$ as 
\begin{eqnarray}
e^{\phi}=e^{\hat x}U^\dagger, \quad 
\left(\hat x={1\over 2}\log(e^\phi e^{\phi ^\dagger })\right).
\end{eqnarray} 
With these two matrices, $S$ is an $N\times N$ matrix and 
is given by the following form 
\begin{eqnarray}
 S^{-1}={\cal U}(y,U) \exp \Big\{ -\psi(y{\bf 1}_N-\hat x/m)
 -\hat x/2 \Big\},
\end{eqnarray}
where ${\cal U}(y,U)$ is an element of the $U(N)$ gauge group satisfying 
\begin{eqnarray}
 {\cal U}(y,U)\rightarrow\left\{
\begin{array}{cc}
 {\bf 1}_N & {\rm for~} y\rightarrow +\infty\\
        U  & {\rm for~} y\rightarrow -\infty\\
\end{array}
\right.
\end{eqnarray}
so that the boundary conditions (\ref{eq:bound_cod}) are satisfied,
and $\psi(y)$ is a certain real smooth function of 
$y$ and satisfies the boundary conditions, 
\begin{eqnarray}
 \psi(y)\rightarrow \left\{
\begin{array}{cc}
  \frac12 my & {\rm for~} y\rightarrow +\infty\\
 -\frac12 my & {\rm for~} y\rightarrow -\infty\\
\end{array}
\right. .
\end{eqnarray}
The BPS equations determine the function $\psi(y)$ uniquely, 
which has been investigated numerically. 

The moduli space of domain walls in 
the GSY model 
is parameterized by $e^{\phi}$
and therefore turns out to be 
\beq
 {\cal M} \simeq GL(N,{\bf C}) [= U(N)^{\bf C}]
\simeq {\bf C}^* \times SL(N,{\bf C}) .
 \label{eq:GSY-moduli}
\eeq
This moduli space admits the isometry 
\beq
 e^{\phi} \to e^{i \alpha} g_{\rm L} e^{\phi} g_{\rm R}^{\dagger} 
  \label{eq:isometry}
\eeq
with $(g_{\rm L},g_{\rm R}) \in SU(N)_{\rm L} \times SU(N)_{\rm R}$ 
and $e^{i \alpha} \in U(1)_{\rm A}$. 
This is because the domain wall solutions 
break the symmetry of the two vacua 
$(N,0)$ and $(0,N)$,  
$G = SU(N)_{\rm C+L} \times SU(N)_{\rm C+R} \times U(1)_{\rm C+L-R}$,  
down to its subgroup. 
The unbroken subgroup is not unique 
as explained in the next subsection.
Here the $y$-dependence of the gauge transformations 
varies for different factors of $G$.  
For instance the gauge transformation $g(y) \in U(N)_{\rm C}$ 
in $SU(N)_{\rm C+L}$ 
has the $y$-dependence such as 
\begin{eqnarray}
 g(y) 
 \rightarrow \left\{
\begin{array}{cc}
 g_{\rm L}^{-1} &  {\rm for~} y\rightarrow \infty\\
              0 &  {\rm for~} y\rightarrow -\infty\\
\end{array}
\right. ,
\end{eqnarray}
with $g_{\rm L} \in SU(N)_{\rm L}$. 
The opposite dependence is for $SU(N)_{\rm C+R}$. 
In the following we do not explicitly write ``C" 
as the indices of the groups.

\subsection{Symmetry structure of the moduli space}\label{subsec:sym}
The $GL(N,{\bf C})$ matrix  
$e^\phi$ can be 
always diagonalized with two unitary matrices
$U_{\rm L},\,U_{\rm R}$ as
\begin{eqnarray}
 e^\phi=U_{\rm L}e^{\phi_0}U_{\rm R}^\dagger,\quad 
\phi_0=m\, {\rm diag}\,(y_1,y_2,\cdots,y_N). 
\label{eq:decomposition}
\end{eqnarray}
These matrices $U_{\rm L},\,U_{\rm R}$ and $\phi_0$ 
give another parametrization of the moduli space 
and are related to $\hat x$ and $U$ as,
\begin{eqnarray}
U=U_{\rm R}U_{\rm L}^\dagger,\qquad 
\hat x=U_{\rm L}\phi_0U_{\rm L}^\dagger.  
\end{eqnarray}
Here the flavor symmetries 
$g_{\rm L}\in SU(N)_{\rm L},\; 
g_{\rm R}\in SU(N)_{\rm R}$
and $e^{-i\alpha}\in U(1)_{\rm A}$ act on $e^\phi$ as
\begin{eqnarray}
 H_0\left(
\begin{array}{cc}
g_{\rm L}e^{\frac{i}2\alpha}&0 \\
0&g_{\rm R}e^{-\frac{i}2\alpha}
\end{array}
\right)=\sqrt{c}\left(g_{\rm L}e^{\frac{i}2\alpha},\,
e^\phi g_{\rm R}e^{-\frac{i}2\alpha}\right) \sim
\sqrt{c}\left({\bf 1}_N,\,
e^{-i\alpha}g_{\rm L}^\dagger e^\phi g_{\rm R}\right), 
\end{eqnarray}
where the last equivalence is due to the $V$-transformation 
(\ref{eq:Vtransformation}). 
Therefore, by using the flavor symmetries, the matrix 
$\phi$ always reduces to the real diagonal matrix $\phi_0$ 
as in Eq.(\ref{eq:decomposition}) 
and each real parameter $y_r$ indicates the position of 
the $r$-th wall.
In this parametrization of the moduli space, 
there is a redundancy such that
\begin{eqnarray}
 U_{\rm L, R}\quad \rightarrow \quad U_{\rm L,R}'=U_{\rm L,R}e^{i \lambda}, \hs{10} e^{i \lambda} \in U(1)^{N},
\label{eq:redundantsym}
\end{eqnarray}
with a real diagonal matrix $\lambda$. 
Furthermore, when some walls are coincident, 
the redundancy is enhanced to a larger group. 
For instance when 
first and second walls are coincident $y_1=y_2$, 
the redundancy is enhanced to $U(1)^{N-1} \times SU(2)$. 
This means that 
$U_L,\,U_R,\,\phi_0$ do not parametrize the moduli 
space correctly 
when some of the walls are coincident. 
Therefore, this parametrization is applicable only for 
separated walls, although physical meanings of the moduli 
parameters are manifest.\footnote{ 
Similar pathology exists in a parametrization of 
the moduli space of the non-Abelian vortices by using 
their position moduli and orientational moduli in the internal space.
In such a parametrization,  
separated vortices are well described but 
coincident vortices cannot be described 
\cite{Eto:2005yh}. 
The smooth coordinates parametrizing the moduli space 
are linear parameters in the moduli matrix 
(see the third reference in \cite{Eto:2006mz}). 
}

In the rest of this subsection we discuss 
the Nambu-Goldstone (NG) modes 
and the quasi-NG modes in our model. 
For a while let us consider the case that 
the ``chiral" symmetry 
$SU(N)_{\rm L} \times SU(N)_{\rm R} \times U(1)_{\rm A}$ 
acts on $e^{\phi}$ as (\ref{eq:isometry}). 
When $\phi$ is eventually proportional to the unit matrix,
the ``chiral" symmetry 
(\ref{eq:isometry}) is spontaneously 
broken down to its diagonal subgroup 
$SU(N)_{\rm V}$ defined by 
$g_{\rm L} =g_{\rm R}$ in (\ref{eq:isometry}):
\beq
 e^{\phi} \to g e^{\phi} g^\dagger, \hs{5} g \in SU(N) .
 \label{eq:SU(N)V}
\eeq
This breaking results in the appearance of 
the massless Nambu-Goldstone bosons (pions),  
parametrizing the coset space 
$[SU(N)_{\rm L} \times SU(N)_{\rm R} \times U(1)_{\rm A}]/ 
SU(N)_{\rm V} \simeq U(N)_{\rm A}$. 
It is known in the supersymmetric case 
that there must appear more massless bosons 
called the quasi-Nambu-Goldstone bosons \cite{Bando:1983ab} 
in order to have K\"ahler target spaces.
In this case we 
can 
have the quasi-NG modes 
as many as NG modes. 
It was found in \cite{Kotcheff:1988ji} 
that the numbers of NG modes and quasi-NG modes 
can change from point to point in the moduli space 
in non-compact nonlinear sigma models, 
although the total number of massless bosons is 
unchanged. 
This is because 
the vacuum expectation values (VEV) along directions 
corresponding to quasi-NG bosons 
can further break the symmetry.
The most general effective K\"ahler potential compatible 
with the symmetry is given in Appendix~\ref{appendix:2} 
to describe the low energy dynamics of massless fields. 
The exchange of NG and quasi-NG modes occurs 
also in the moduli space of multiple non-Abelian vortices \cite{Eto:2004ii}.

Note the fact that 
the global symmetry 
$G = SU(N)_{\rm L} \times SU(N)_{\rm R} \times U(1)_{\rm A}$ 
in (\ref{eq:isometry}) 
acts on the moduli space metric as an isometry 
whereas the complexified group 
$G^{\bf C} =SL(N,{\bf C})_{\rm L} 
\times SL(N,{\bf C})_{\rm R} \times {\bf C}^*$ 
acts on 
it transitively but not as an isometry.
Therefore $G^{\bf C}$ 
action may change the point in moduli space to 
another with a different symmetry structure. 
By using the $G^{\bf C}$ action, 
arbitrary moduli parameter 
$\phi$ can be brought to zero: 
\beq 
 e^{\phi} = {\bf 1}_N. 
 \label{eq:max-vac}
\eeq
At this point in moduli space, 
the global symmetry $G$ is broken down to 
$H_{\rm max} = SU(N)_{\rm V}$ defined in (\ref{eq:SU(N)V}).
Then the number of the NG modes is $\dim G/H_{\rm max} = N^2$. 
Since the total number of massless bosons is 
$\dim G^{\bf C}/H^{\bf C} = 2N^2$, 
the number of quasi-NG modes\footnote{ 
This situation that the number of the NG bosons and 
quasi-NG bosons coincide 
is called maximal realizations \cite{Bando:1983ab} 
or fully doubled realizations \cite{Kotcheff:1988ji}.
} is  
$2N^2 - N^2 = N^2$ at this point of moduli space.  

Since the symmetry of Lagrangian is $G$ but not $G^{\bf C}$ 
we can use only $G$ when we discuss the symmetry 
structure at each point in moduli space. 
General $\phi$ can be transformed by $G$ to 
\beq
 e^{\phi} = {\rm diag.} (v_1, v_2, \cdots, v_N) 
 \label{eq:general-vac}
\eeq
with $v_i$ real parameters.
When all $v_i$'s are different from each other, 
$H_{\rm max} = SU(N)_{\rm V}$ is further broken down 
to 
$H_{\rm min} = U(1)_{\rm V}^{N-1}$.
Here 
the numbers of NG bosons and quasi-NG bosons
are $2N^2 - (N -1)$ and 
$N-1$, respectively. 
These $N-1$ quasi-NG bosons correspond to 
the $N-1$ parameters $v_i$ without the overall factor.
Therefore some of quasi-NG bosons 
at the symmetric point 
(\ref{eq:max-vac}) 
in the moduli space 
change to the NG bosons 
parametrizing
$H_{\rm max}/H_{\rm min} 
= SU(N)_{\rm V}/U(1)_{\rm V}^{N-1}$ 
reflecting this 
further symmetry breaking. 
When some $v_i$'s coincide, 
some non-Abelian groups 
are recovered: 
$H = U(1)_{\rm V}^r \times \prod U(n_i)_{\rm V}$.
Then the NG modes $H_{\rm max}/H$ are 
supplied from quasi-NG modes.\footnote{
The space $H_{\rm max}/H_{\rm min}$ or $H_{\rm max}/H$ 
is fibered over $G/H_{\rm max}$ and 
the total space of NG bosons is of course 
$G/H_{\rm min}$ or $G/H$. 
These spaces are $G$-orbits in the full moduli space 
${\cal M} \simeq GL(N,{\bf C})$, 
and the latter is stratified by 
these spaces as leaves. 
}
All these points in the moduli space 
with different unbroken symmetries are of course degenerate. 
This ``vacuum alignment'' was first pointed out by 
G.~Shore \cite{Kotcheff:1988ji} in the context of supersymmetric 
nonlinear sigma models.

An interesting point is that the 
diagonal moduli parameters $v_i$ 
(quasi-NG bosons)
in Eq.~(\ref{eq:general-vac}) 
correspond to the positions of $N$ domain walls, 
see Eq.~(\ref{eq:decomposition}).
When all domain walls are separated, 
the unbroken symmetry 
is $U(1)_{\rm V}^{N-1}$. 
When positions of $n$ domain walls coincide, $U(n)_{\rm V}$ 
symmetry is recovered. 
This phenomenon 
has a resemblance to 
the case of D-branes.  
However, there is a crucial difference: 
the symmetry in our case of domain walls is a global 
symmetry, whereas that of D-branes is a local 
gauge symmetry. 
However in the case of the $d=2+1$ wall world-volume, 
massless scalars can be dualized to gauge fields.
Shifman and Yung \cite{Shifman:2003uh} expected that 
the off-diagonal gauge bosons of $U(N)$ 
(which are originally the off-diagonal NG bosons of $U(N)$
before taking a duality) 
will become massive when domain walls are separated,   
in order to interpret domain walls as 
D-branes. 
However, our analysis shows that 
the off-diagonal NG bosons of $U(N)$
remain massless, and 
instead 
some of the quasi-NG bosons 
become NG bosons for further symmetry breaking 
with the total number of massless bosons unchanged  
as explained. 
We will take a duality explicitly in Sec.~\ref{sc:duality} 
in the case that the dimension of the wall world volume 
is $3+1$.

\subsection{The Effective action of domain walls}\label{subsec:efac-GSY}
Elements of the matrix $\phi$ are 
holomorphic coordinates of the moduli space. 
In the effective action, the matrix $\phi$ is promoted 
to a matrix-valued field. 
Note that matrix-valued fields $U$ and $\hat x$ 
($U_{\rm R},\,U_{\rm L}$ and $\phi_0$) depend on both 
$\phi$ and $\phi^*$ (neither holomorphic nor 
anti-holomorphic with respect to $\phi$). 
With these knowledge, the K\"ahler potential for the 
effective action is calculated by the formulas 
(\ref{eq:formal-Kahler}) and 
(\ref{eq:formal-Kahler-density0}), 
where  $\Omega$ and $\Omega_0$ are given by
\begin{eqnarray}
 \Omega
= \exp \Big \{2\psi (y{\bf 1}_N-\hat x/m)+\hat x \Big\},
\quad
\Omega_0 
= e^{\hat x} \Big(e^{m(y{\bf 1}_N-\hat x/m)}
+e^{-m(y{\bf 1}_N-\hat x/m)} \Big).
\end{eqnarray}
If we abandon to obtain the density ${\cal K}_{ij^*}$ 
of the K\"ahler metric, 
we can calculate the K\"ahler metric directly 
without any approximations. 
The formulas 
(\ref{eq:formal-Kahler})--(\ref{eq:formal-Kahler-density0}) 
tell us that the quantity $\hat x$ in 
the K\"ahler potential is the only matrix which is 
not proportional to the unit matrix. 
Moreover the matrix valued fields $\phi$ and $\phi^\dagger$
appear only through the matrix $\hat x$. 
Therefore we can write the K\"ahler potential 
in terms of a function $F$ of the matrix $\hat x$ as 
\begin{eqnarray}
 K(\phi,\phi^\dagger) = {\rm Tr}[F(\hat x)].
\end{eqnarray}
This result reflects the fact that 
if the matrix $\phi$ (thus $\hat x$) is diagonal,
the solution reduces to a direct sum of the solutions for 
independent walls. 
Because of the K\"ahler invariance, the K\"ahler metric 
receives no contribution from purely holomorphic or purely 
anti-holomorphic additive terms in the K\"ahler potential. 
This fact implies that the function $F(x)$ is equivalent 
under arbitrary linear transformations,  
\begin{eqnarray}
F(x) \simeq F(x)+a x+b,
\end{eqnarray}
since ${\rm Tr} (\hat x)$ can be written as a 
sum of holomorphic and anti-holomorphic functions 
\begin{eqnarray}
 2 \, {\rm Tr}(\hat x)
=\log {\rm det}(e^\phi e^{\phi^\dagger})
=\log {\rm det}(e^\phi) + \log {\rm det}(e^{\phi^\dagger})
={\rm Tr}(\phi +\phi ^\dagger). 
\end{eqnarray}
Actually, the K\"ahler potential in Eq.(\ref{eq:formal-Kahler}) 
is well-defined only after using this K\"ahler transformation, 
since it contains divergent parts due to 
constant terms and linear terms with respect to $\hat x$. 
Since the function $F$ is independent of the size of matrix, 
$N$, the function $F$ can be determined 
by considering the Abelian case ($N=1$). 
In the Abelian case, the complex field $\phi$ consists of two 
real fields corresponding to two Nambu-Goldstone modes: 
${\rm Re}\,\phi/m$ and ${\rm Im}\,\phi$, which are 
the Nambu-Goldstone modes 
for broken translation and $U(1)$ phase, respectively. 
The low energy theorem 
(\ref{eq:inertial-mass}) 
for these Nambu-Goldstone modes 
tells 
us that the K\"ahler potential for $N=1$ is given by $K = c \, \hat x^2/m$.
Thus we obtain the K\"ahler 
potential for general $N$ in a compact form 
\begin{eqnarray}
 K(\phi ,\phi ^\dagger)
= \frac{c}{m}{\rm Tr}[\hat x^2]
={c\over 4m}{\rm Tr}\left[(\log e^\phi e^{\phi ^\dagger})^2\right].
 \label{eq:compact-kahler}
\end{eqnarray}
This is a K\"ahler potential on 
${\cal M} \simeq GL(N,{\bf C})$.

Next let us derive the K\"ahler metric from this K\"ahler potential.
To this end, 
it is convenient to define derivative operators $\delta_\mu$ and $\delta_\mu^\dagger$ such that 
$\delta_\mu \equiv \p_\mu \phi \frac{\p}{\p \phi},~\delta_\mu^\dagger \equiv \p_\mu \phi^\dagger \frac{\p}{\p \phi^\dagger}$.
For instance, $\delta_\mu$ acts on $\hat x$ as,\footnote{
The relation between infinitesimal deformations 
$\delta e^{2\hat x}$ 
and $\delta \hat x$ 
is generally given by  
\begin{eqnarray}
\delta e^{2\hat x}e^{-2\hat x}=
2 \int^1_0 dt \, \left(e^{2t \hat x} \, 
\delta \hat x \, e^{2(1-t)\hat x} \right) e^{-2\hat x} 
= 2 \int^1_0 dt \, e^{2t L_{\hat x}} \times \delta \hat x =
\frac{e^{2L_{\hat x}}-1}{L_{\hat x}} \times \delta \hat x.
\end{eqnarray}
}  
\begin{eqnarray}
2 \, \delta_\mu \hat x
=\frac{2L_{\hat x}}{e^{2L_{\hat x}}-1} \times \pi_\mu =
\pi_\mu - [\hat x,\, \pi_\mu] 
+ \frac{1}{3} [\hat x,[\hat x, \pi_\mu]] + \cdots , 
\end{eqnarray} 
where $\pi_\mu$ is defined by 
$\pi_\mu\equiv (\partial_\mu e^{\phi})e^{-\phi}$,
and $L_{\hat x}$ is an anti-Hermitian operator acting 
as $L_V\times A=[V,\,A]$.
The effective Lagrangian is, thus, calculated as
\begin{eqnarray}
{\cal L}=  \delta^\mu\delta_\mu^\dagger K(\phi,\phi^\dagger)
=\frac{c}{2m}{\rm Tr}
\left[\pi_\mu^\dagger\,\frac{2L_{\hat x}}{e^{2L_{\hat x}}-1}
\times \pi^\mu\right]. \label{eq:compact-Kahler-pot} 
\end{eqnarray}
Here we have used the identity 
$2 \, {\rm Tr}[\hat x \, \delta_\mu^\dagger \hat x] 
= {\rm Tr}[\hat x\,\pi_\mu^\dagger]$.

\subsection{Localization properties in strong coupling limit}
\label{subsec:localization}
Here, we examine the localization properties of various 
massless modes. 
We will use the density of the K\"ahler metric 
or K\"ahler potential as physical quantities to examine 
the localization properties of massless modes. 

To this goal, it is convenient to
consider the strong coupling limit $g^2\rightarrow \infty$ 
where 
we know the exact solution for the matrix valued function 
$\Omega$ which is given in terms of the moduli matrix $H_0$ as 
\begin{eqnarray}
\Omega =\Omega_0 =c^{-1} H_0 e^{2My} H_0^\dagger 
= e^{\hat x} \Big( e^{m(y {\bf 1}_N - \hat x/m)} 
+ e^{-m(y {\bf 1}_N - \hat x/m)} \Big). 
\end{eqnarray} 
Eq.(\ref{eq:formal-Kahler-density0}) 
gives the density of the K\"ahler metric 
in the strong coupling limit 
as 
\begin{eqnarray}
\delta^\mu\delta_\mu^\dagger {\cal K}(y,\phi,\phi^\dagger)=
\delta^\mu\delta_\mu^\dagger {\rm Tr}\left[c\,\log\Omega\right]
&=& c \, {\rm Tr}\left[\pi_\mu^\dagger \Omega^{-1}
\pi^\mu \Omega^{-1}e^{2\hat x}\right].
\label{eq:kahler-metric-density}
\end{eqnarray} 
By integrating over $y$ one can easily check that this 
density of the K\"ahler metric leads to the effective 
Lagrangian (\ref{eq:compact-Kahler-pot}).
Let us introduce an $N\times N$ matrix $\tau_\mu$ as
\begin{eqnarray}
\tau_\mu &\equiv& U_{\rm L}^\dagger 
\left(2(e^{L_{\hat x}}+1)^{-1} \times \pi_\mu \right)U_{\rm L}
\nn\\
&=&\partial_\mu \phi_0+2(e^{L_{\phi_0}}+1)^{-1} 
\times \left(U_{\rm L}^\dagger \partial_\mu U_{\rm L} \right) 
- 2(e^{-L_{\phi_0}}+1)^{-1}\times 
\left(U_{\rm R}^\dagger \partial_\mu U_{\rm R}\right).
\end{eqnarray}
If $r$-th and $s$-th walls are well-separated $y_r \gg y_s$, 
the $(r,s)$ component of the matrix $\tau_\mu$ is given by 
\begin{eqnarray}
 (\tau_\mu)_{rs} \approx \left\{
\begin{array}{cc}
-(U_{\rm R}^\dagger \partial_\mu U_{\rm R})_{rs}  & {\rm for~} r>s,\\
 m\partial_\mu y_r+(U_{\rm L}^\dagger \partial_\mu U_{\rm L})_{rr}-
(U_{\rm R}^\dagger \partial_\mu U_{\rm R})_{rr}  & {\rm for~} r=s, \\
(U_{\rm L}^\dagger \partial_\mu U_{\rm L})_{rs}  & {\rm for~} r<s.
\end{array}
\right.
\end{eqnarray}
In term of this $\tau_\mu $, 
the density of the K\"ahler metric is given by
\begin{eqnarray}
{\cal K}_{i j^*}\partial _\mu \phi ^i\partial ^\mu \phi ^{j*}
= \frac{c}4\sum_r^N {|(\tau_\mu)_{rr}|^2\over \cosh^2(m(y-y_r))} 
+ c\,\sum_{r\not=s}^N{\cosh^2\left(\frac{m}2(y_r-y_s)\right)|
(\tau_\mu)_{rs}|^2 \over \cosh(m(y-y_r))
\cosh(m(y-y_s))}.
\label{eq:density-Kahler2}
\end{eqnarray}
This formula contains full information of the localization 
properties of the massless modes.

Eq.~(\ref{eq:density-Kahler2}) shows that the fields 
$y_r$ indicates that the fluctuation field of
the position of the $r$-th wall and the wave function 
corresponding to the $r$-th diagonal element 
$(\tau_\mu)_{rr}$ is localized on the $r$-th wall. 
On the other hand, the fluctuation modes of the off-diagonal 
elements $(\tau_\mu)_{rs},(r\not =s)$ are not localized 
on the individual wall. 
To see where these off-diagonal modes have non-vanishing 
wave functions, we take the limit of well-separated walls 
$y_r \gg y_s$. 
Then we obtain 
\begin{eqnarray}
&&{\cosh^2\left(\frac{m}2(y_r-y_s)\right)
\over \cosh(m(y-y_r))\cosh(m(y-y_s))}
\approx \left\{
\begin{array}{ccc}
  0,& {\rm for~}y\gg y_r\\
 1, &{\rm for~} y_s\ll y \ll y_r\\
0,&{\rm for~} y\ll y_s
\end{array}\right. .
\end{eqnarray}
Therefore we find that the off-diagonal elements 
$(\tau_\mu)_{rs}$ correspond to the non-Abelian clouds 
which have support between the $r$-th wall and the $s$-th wall. 
As we showed in the previous section, 
non-vanishing fluctuation of these modes cause a 
repulsive force between the two walls.  
In contrast, the bulk modes have support over the entire 
space including infinity, and the localized modes have 
support between (possibly coincident) walls.  
Note that $4k(N-k)$ modes of the non-Abelian clouds, 
which correspond to $(\tau_{\mu})_{rs}$ and 
$(\tau_\mu)_{sr}$ with $1\le r\le k$ and 
$k+1 \le s \le N$, have support in the $k$-th vacua 
in Eq.(\ref{eq:kthvacuum}) 
and just constitutes the NG modes of 
that vacua.

\subsection{Dynamics of non-Abelian cloud fluctuations: chiral dynamics}
\label{subsec:dynamics-GSY}
If we restrict our attention to coincident walls 
$y_1=y_2=\cdots=y_N=0$, the matrix $e^\phi$ reduces to a 
unitary matrix $U^\dagger=U_{\rm L}U_{\rm R}^\dagger$ 
leading to 
$\hat x=0$, then the Lagrangian reduces to 
the chiral Lagrangian plus
a kinetic term for fluctuations of $\hat x$ as 
\begin{eqnarray}
 {\cal L} 
= -\frac{c}{2m}{\rm Tr}\left[U^{\dagger} \partial_\mu U \, 
U^{\dagger} \partial^\mu U\right]
+\frac{c}{2m}{\rm Tr}\left[\partial_\mu \hat x \, 
\partial^\mu\hat x \right]
 + {\cal O}(\hat x^4).
\label{eq:dressed-chiral}
\end{eqnarray}
This is nothing but the chiral Lagrangian for 
the chiral symmetry breaking 
if we set all the quasi-NG bosons to zero; $\hat x=0$. 
There NG bosons are interpreted as 
``pions''. 
In Ref.~\cite{Eto:2005cc}
we placed two domain walls at the same position 
in order to realize the chiral Lagrangian.

Conversely, in the case of well-separated walls,  
$y_1\gg y_2\gg \cdots \gg y_N$, 
the Lagrangian asymptotically reduces to
\begin{eqnarray}
 {\cal L}\big|_{\hbox{\tiny well-separated}}&\approx&
\sum_{r}\frac{cm}{2}|\partial_\mu y_r |^2
+\frac{c}{2m}\left|(A^-_\mu)_{rr}\right|^2\nonumber\\
&&+\frac{c}{4}\sum_{r\not =s}|y_r-y_s|\left(|(A_\mu^+)_{rs}|^2+
|(A_\mu^-)_{rs}|^2\right),
\label{eq:AsymLag}
\end{eqnarray}
where the vector fields $A_\mu^\pm$ give the fluctuations 
of the unitary matrices $U_{\rm L}$ and $U_{\rm R}$ as
\begin{eqnarray}
 A_\mu^{-}&=&iU^\dagger_{\rm L}\partial_\mu U_{\rm L}- 
iU^\dagger_{\rm R}\partial_\mu U_{\rm R}=
-U_{\rm L}^\dagger\left(U^\dagger i\partial_\mu U\right) U_{\rm L}\nn,\\
 A_\mu^{+}&=&iU^\dagger_{\rm L}\partial_\mu U_{\rm L} +
iU^\dagger_{\rm R}\partial_\mu U_{\rm R}.
\end{eqnarray}
Due to the redundancy (\ref{eq:redundantsym}),
the diagonal elements of $A_\mu^+$ 
turn out to be unphysical modes as we observe in (\ref{eq:AsymLag}).
Note that kinetic terms for the off-diagonal elements of 
 $A_\mu^{\pm}$ 
are proportional to the distance of walls $|y_r-y_s|$.
This fact tells us that these are non-Abelian clouds as we expected. 

Let us consider  well-separated domain walls (kinks) 
in the case of $N=2$ for simplicity.
We again ignore the center of mass position and 
the Nambu-Goldstone mode for broken overall $U(1)$ phase.
We parametrize $U_{\rm L},\,U_{\rm R} \in {SU(2)}$ as
\begin{eqnarray}
 U_{\rm L}= \left(
\begin{array}{cc}
 \displaystyle \cos \left(\frac{\theta_L}2\right) 
\exp \left( i\frac{\varphi_L+\xi_L}2 \right) 
&  \displaystyle \sin \left(\frac{\theta_L}2\right) 
\exp \left(i\frac{\varphi_L-\xi_L}2 \right) \\
 \displaystyle -\sin \left(\frac{\theta_L}2\right) 
\exp \left( -i\frac{\varphi_L-\xi_L}2 \right) 
& \displaystyle \cos\left(\frac{\theta_L}2\right) 
\exp \left(-i\frac{\varphi_L + \xi_L}2 \right) 
\end{array}\right),
\end{eqnarray} 
and similarly for $U_{\rm R} \in SU(2)$.
Then the Lagrangian for well-separated walls 
$R=y_1-y_2\gg 1/m$ reduces to
\begin{eqnarray}
{\cal L}\big|^{N=2}_{\hbox{\tiny well-separated}} 
&\approx& \frac{c}{4m^2} \Big[ m^2 (\partial_\mu R)^2 
+ \big( \partial_\mu \xi + \cos \theta_{\rm L} \, 
\partial_\mu \varphi_{\rm L} - \cos \theta_{\rm R} \, 
\partial_\mu \varphi_{\rm L} \big)^2 \Big] \notag \\
&{}& +~ \frac{c R}{4} \Big[ (\partial_\mu \theta_{\rm L})^2 
+ \sin^2 \theta_{\rm L} (\partial_\mu \varphi_{\rm L} )^2 
+ (\partial_\mu \theta_{\rm R})^2 
+ \sin^2 \theta_{\rm R} (\partial_\mu \varphi_{\rm R} )^2 \Big], 
\end{eqnarray}
where $\xi \equiv \xi_{\rm L} - \xi_{\rm R}$. 
The mode $\xi_{\rm L}+\xi_{\rm R}$ is unphysical and does not 
appear in this effective Lagrangian. 
As in the case of the walls discussed in Section 3, 
the fields $\theta_{\rm L,R}$ and $\varphi_{\rm L,R}$ 
have kinetic terms whose coefficients are proportional to 
the distance of the walls $R$ for large $R$. 
They correspond to the non-Abelian clouds which parametrize 
the vacuum between the walls 
${\bf C}P^1\times {\bf C}P^1\simeq S^2\times S^2$.
Therefore, the conserved charges for the non-Abelian 
clouds lead to a long-range repulsive force as in Sec.~\ref{section:3}.

On the other hand, 
the addition of small imaginary masses for the Higgs fields 
leads a long-range attractive force. 
For instance, let us consider a deformation of the mass matrix
\begin{eqnarray}
 M=\frac12{\rm diag}\,(m+i\tilde m_{\rm L},\, 
m-i\tilde m_{\rm L},\, -m+i\tilde m_{\rm R},\, 
-m-i\tilde m_{\rm R})
\end{eqnarray}
with small mass parameters $\tilde m_{\rm L,R} \ll m$. 
These mass parameters $\tilde m_{\rm L,R}$ break the 
chiral symmetry 
$SU(2)_{\rm L} \times SU(2)_{\rm R}$ to 
$U(1)_{\rm L} \times U(1)_{\rm R}$ and induce a 
long-range attractive potential through Killing vectors 
$k_{\rm L,R} = \tilde m_{\rm L,R} \frac{\p}{\p \varphi_{\rm L,R}}$. 
With very small charges $Q\ll c\frac{\tilde m_{\rm L,R}}m(\ll c)$, 
an expectation value of $R$ is guessed to be small,
$R \approx Q/(c \,\tilde m_{\rm L,R})\ll 1/m$. 
In the low energy limit, $E\gtrsim Q \tilde m\rightarrow 0$, 
we obtain the dressed chiral Lagrangian (\ref{eq:dressed-chiral}) 
with a potential made of the sum of the squares of 
the Killing vectors 
$k_{\rm L} = \frac{\tilde m_{\rm L}}{2} 
(\sigma_3 U)_{ij} \frac{\p}{\p U_{ij}} 
+ i \frac{\tilde m_{\rm L}}{2} 
[\sigma_3,\,\hat x]_{ij} \frac{\p}{\p \hat x_{ij}}$ 
and $k_{\rm R} = - \frac{\tilde m_{\rm R}}{2} 
(U \sigma_3)_{ij} \frac{\p}{\p U_{ij}}$:  
\begin{eqnarray}
V=\frac{c}{4m}\left(\tilde m_{\rm L}^2+\tilde m_{\rm R}^2
-\tilde m_{\rm L}\tilde m_{\rm R}{\rm Tr}[U^\dagger \sigma_3U\sigma_3]+
\frac{\tilde m_{\rm L}^2}2{\rm Tr} \, [i\sigma_3,\,\hat x]^2 \right)
.
\end{eqnarray} 
Most quasi-NG bosons become massive by the third term 
while 
the quasi-NG boson corresponding to $\hat x$ 
commuting with $\sigma_3$ remains massless.

We have 
obtained a 
mass term for pions in the second term. 
However, it does not agree with the usual form 
induced by the quark mass terms in the chiral 
perturbation theory. 
The same situation occurs 
in the context of the holographic QCD \cite{Hashimoto:2007fa}.

\section{Duality and Non-Abelian Two-Form Fields}\label{sc:duality}

Up to here we did not restrict 
the dimensionality of the space-time; 
BPS domain walls can be constructed 
in dimensions ranging from 
$d=1+1$ to $d=4+1$. 
In this section we restrict the dimension 
to be the maximal one $d=4+1$ to discuss 
the duality on the 3+1 dimensional 
world-volume of walls, which is realistic for brane-world 
applications. 
In 3+1 dimensions, scalar fields are dual to 2-form fields. 
In the framework of supersymmetry with four supercharges, 
the chiral superfields 
$\Phi(x,\theta,\bar\theta)$ ($\bar D_{\dot\alpha} \Phi = 0$) are dual to 
the chiral spinor superfields $B_{\alpha} (x,\theta,\bar\theta)$
($\bar D_{\dot\alpha} B_{\beta} = 0$) \cite{Siegel:1979ai}.

In the simplest model considered in Sec.~\ref{section:3}, 
the moduli space of domain walls is toric K\"ahler, 
namely it admits $U(1)^n$ holomorphic isometry 
with $n$ its complex dimension. 
In this case the dual theory can be obtained 
by using the $n$ Abelian dualities along 
$n$ $U(1)$ isometries \cite{Lindstrom:1983rt}. 
The dual theory is an interacting 
theory of Abelian 2-form fields.

In this section we discuss the dual theory 
of the GSY model considered in Sec.~\ref{sc:GSY}. 
In the paper of Shifman and Yung \cite{Shifman:2003uh}, 
they considered the $d=3+1$ bulk dimension  
and so the $2+1$ dimensional wall world-volume. 
They claimed that the dual theory of the $U(2)$ NG bosons 
in $2+1$ dimensions is $U(2)$ gauge theory, 
although they were not able to obtain 
non-trivial interaction term of non-Abelian gauge fields. 
Here we construct the full dual theory 
by restricting 
the bulk dimension to $d=4+1$  
so that the wall world-volume has $3+1$ dimensions. 
We thoroughly perform the duality transformation of 
the $GL(N,{\bf C})$ sigma model and 
find the action of non-Abelian two-form fields. 
Its bosonic counterpart is known as 
the Freedman-Townsend model \cite{Freedman:1980us}.
In fact the K\"ahler potential (\ref{eq:compact-kahler}) 
precisely coincides with the one proposed 
for supersymmetric extension of the Freedman-Townsend model
\cite{Clark:1988gx}.

We start from the Lagrangian of the 2-form fields. 
Here we use the superfield formalism 
basically following the notation in \cite{WB}.
The 2-form field $B_{\mu\nu}(x)$ in 
$3+1$ dimensions 
belong to the (anti-)chiral spinor superfields 
$B_{\al}(x,\theta,\thb)$ [$\bar B_{\dot\al}(x,\theta,\thb)$], 
satisfying the constraints~\cite{Siegel:1979ai,Gates:1983nr}
\beq
 \bar D_{\dot\al} B_{\be}(x,\theta,\thb) = 0, \hs{10} 
 D_{\al} \bar B_{\dot\be}(x,\theta,\thb) = 0.
\eeq
These superfields can be expanded 
in terms of component fields as 
\beq
&& \hs{-5} 
 B^{\al}(y,\theta) 
 = \psi^{\al}(y) + \1{2} \theta^{\al}(C(y) + i D(y)) 
  + \1{2} (\sig^{\mu\nu})^{\al\be} \theta_{\be} B_{\mu\nu}(y)
  + \theta\theta
\eta^{\al}(y), \non
&& \hs{-5} 
 \bar B_{\dot\al}(y\dagg,\bar\theta) 
 = \psb_{\dot\al}(y\dagg) 
  + \1{2} \thb_{\dot\al}(C(y\dagg) - i D(y\dagg)) 
  +\1{2} (\bar\sig^{\mu\nu})_{\dot\al\dot\be} \thb^{\dot\be} 
    B_{\mu\nu}(y\dagg)
  + \thb\thb\bar \eta_{\dot\al}(y\dagg), \hs{10}
\eeq
where $(\sig^{\mu\nu})^{\al}{}_{\beta} 
= \1{4} (\sig^{\mu}\sigb^{\nu} 
- \sig^{\nu}\sigb^{\mu})^{\al}{}_{\beta}$, 
$y^{\mu} \equiv x^{\mu} + i\theta\sig^{\mu} \thb$ 
and 
$y^{\mu\dagger} = x^{\mu} - i\theta\sig^{\mu} \thb$. 
If one fixes $y^\mu$ ($y^{\mu\dagger}$), one finds 
$\bar D_{\dot\al} = \del/\del \bar\theta^{\dot\al}$ 
($D_{\al} = - \del/\del\theta^{\al}$). 
See Ref.~\cite{WB} for details. 
We consider the non-Abelian 2-form field 
with the group $G=U(N)$:
$B_{\al}(x,\theta,\thb)=B^A_{\al}(x,\theta,\thb) T_A$. 
Let us introduce a ${\cal U}(N)$-valued auxiliary vector 
superfield 
$A(x,\theta,\thb) = A^A(x,\theta,\thb) T_A$, 
satisfying the constraint ${A^A}\dagg = A^A$. 
Its field strengths are (anti-)chiral spinor superfields,  
\beq
 W_{\al} = - \1{4 } \bar D \bar D 
             (e^{-A}D_{\al}e^{A}) ,\hs{10}
 \bar W_{\dot\al} 
  = \1{4 } D D (e^{A} \bar D_{\dot\al}e^{-A}). 
 \label{Auxiliary-fs}
\eeq
The first-order Lagrangian is given as \cite{Clark:1988gx}
\beq
 {\cal L} = - \1{2f} \left[\int d^2\theta\; {\rm Tr} (W^{\al} B_{\al}) 
 + \int d^2 \thb \; {\rm Tr} (\bar W_{\dot\al} \bar B^{\dot\al}) \right]
 + \1{4f} \int d^4 \theta\; {\rm Tr} A^2. \label{Lag.}
\eeq
See also \cite{Lindstrom:1983rt} for the Abelian case.
This Lagrangian 
is invariant under the anti-symmetric tensor 
gauge 
transformation\footnote{
This transformation is Abelian, 
though $\Omega$ is ${\cal G}$-valued.
}, 
parameterized by a ${\cal U}(N)$-valued vector superfield 
$\Omega(x,\theta,\thb)=\Omega^A(x,\theta,\thb)T_A$ 
(with ${\Omega^A}\dagg =\Omega^A$):
\beq
 && \delta B_{\al} = - \2{4} \bar D \bar D 
    {\cal D}_{\al} (e^{-A}\Omega), \hs{10} 
 \delta \bar B_{\dot\al} 
 = - \2{4} D D \bar {\cal D}_{\dot\al} (\Omega e^{-A}),\non
 && \delta A = 0  \label{SUSY-TGT}
\eeq
with the covariant spinor 
derivative 
${\cal D}_{\al} = D_{\al} 
+ [e^{- A} D_{\al}e^{A}, \,\cdot\,]$.
With this invariance we can take the Wess-Zumino gauge: 
$D=\psi_{\alpha}=0$.
The Lagrangian (\ref{Lag.}) is invariant under 
the {\it global} $U(N)$-transformation
\beq
 &&B_{\al} \to B_{\al}' = g^{-1} B_{\al} g, \hs{10} 
   \bar B_{\dot\al} \to \bar B_{\dot\al}' 
   = g^{-1} \bar B_{\dot\al} g, \non 
 && A \to A' = g^{-1} A g , \hs{10} 
    W_{\al} \to W_{\al}' = g^{-1} W_{\al} g , 
 \label{G-action}
\eeq
with $g \in U(N)$.

In principle the second order Lagrangian 
of the 2-form fields $B_{\alpha}$
can be obtained 
by eliminating the auxiliary field $A$ with
solving its equations of motion.
On the other hand, if we eliminate $B_{\al}(x,\theta,\thb)$,  
we can obtain the $GL(N,{\bf C})$ sigma model.
The equation of motion for $B_{\al}$
\beq
 -4 W_{\al}(x,\theta,\thb) 
  = \bar D \bar D (e^{-A}D_{\al}e^{A}) 
  = 0  \label{vanish}
\eeq
implies that $A$ is in a pure gauge:
\beq
 e^{A(x,\theta,\thb)} 
 = e^{\phi(x,\theta,\thb)} e^{\phi\dagg(x,\theta,\thb)}, \hs{10} 
 \bar D_{\dot\al} \phi(x,\theta,\thb) = 0 .  \label{pure_gauge}
\eeq
Here we have introduced the ${\cal U}(N)$-valued chiral 
superfield $\phi=\phi^A T_A$.
By substituting (\ref{pure_gauge}) back into the Lagrangian (\ref{Lag.}), 
we obtain the Lagrangian for $\phi$ 
\cite{Clark:1988gx}: 
\beq
 {\cal L} 
 = \int d^4\theta\; \1{4f} 
 {\rm Tr} \left[(\log e^{\phi} e^{\phi\dagg})^2\right]. 
 \label{nlsm_on_GC}
\eeq
This coincides with the K\"ahler potential (\ref{eq:compact-kahler})
with identifying $c/m = 1/f$.
In the Wess-Zumino gauge $D=\psi_{\alpha}=0$ 
physical bosonic fields are
the 2-form fields $B_{\mu\nu}^A$ and 
the associated scalar fields $C^A$. 
When all domain walls are coincident 
we can identify the $B_{\mu\nu}^A$ as the NG bosons of $U(N)$
and $C^A$ as the quasi-NG bosons. 
When some walls are separated, identification is rather complicated.

\section{A Comment on Non-Abelian Monopoles 
and a Monopole Bound State}\label{sc:monopole}

Our work is straightforwardly applicable 
to a system of confined monopoles in the Higgs phase. 
Those monopoles can be identified with kinks 
inside a non-Abelian vortex \cite{Tong:2003pz}. 
It is well known that a single BPS vortex in $U(1)\times SU(N)$ gauge theory
coupled to $N$ Higgs fields in the fundamental representation 
with the FI term has 
the orientational moduli $\vec \phi\in {\bf C}P^{N-1}$ 
($\vec \phi\simeq \lambda \vec \phi$ with $\lambda \in {\bf C^*}$) 
\cite{Hanany:2003hp,Auzzi:2003fs}. 
Its K\"ahler potential is given by 
\begin{eqnarray}
 K=\frac{4\pi}{g^2}\log|\vec \phi|^2\label{eq:Kahler-orientation}
\end{eqnarray}
with $g$ the coupling constant of $SU(N)$.
If we add real masses described by a diagonal mass matrix $M$ 
in the original theory, 
a contribution to the effective theory on the vortex is calculated 
by a Killing vector $\delta \vec \phi=iM \vec \phi$; 
The potential is written as the square of the Killing vector 
\cite{Tong:2003pz,Hanany:2004ea,Shifman:2004dr,Eto:2004rz}. 
This system is just the same as 
the one which we considered in 
the strong gauge coupling limit
in Sec.~\ref{section:3}, if we replaced $4\pi /g^2$ by $c$.
Here, kinks (domain walls) of the effective action 
correspond to monopoles confined by vortices attached 
from both sides.
Actually, the coefficient in the potential (\ref{eq:Kahler-orientation}) 
can be determined so that the tension of the kink 
coincides with the mass of the monopole 
\cite{Hanany:2004ea, Eto:2004rz}. 
So far only non-degenerate masses $M$ were considered 
for the Higgs scalar fields
\cite{Tong:2003pz,Hanany:2004ea,Shifman:2004dr,Eto:2004rz}. 
In this case a confined monopole is Abelian 
(of the 't Hooft-Polyakov type) 
and attached vortices are also Abelian (of the ANO type).  
A new aspect in this paper is that 
if we choose the degenerate masses $M$ as discussed in 
Sec.~\ref{section:3}, 
non-Abelian monopoles are confined by non-Abelian vortices. 
This precisely gives a correspondence between 
non-Abelian domain walls and non-Abelian monopoles as 
discussed in Introduction.
In particular, we expect correspondence of 
non-Abelian clouds in both solitons. 
We expect that in the original theory 
we can take a limit 
of usual non-Abelian monopoles 
without vortices in an unbroken phase
by 
turning off the FI parameter.  
In this limit the $U(1)$ magnetic flux
spreads out and the vortex vanishes.
This is  
because the K\"ahler potential 
is independent of the FI parameter and the $U(1)$ 
gauge coupling.
More precise correspondence to 
non-Abelian monopoles 
deserves to be studied further 
in particular for the application to non-Abelian duality.

An interesting application of this correspondence is 
a monopole-monopole bound state.
A mass splitting in the imaginary part between
masses, which are degenerate in the real part, 
can be considered by taking another Killing vector; 
If we take masses like Eq.~(\ref{eq:additional-masses}),
we see the existence of 
the long-range repulsive force by charges $Q$ and 
the confining force by imaginary masses $\tilde m$
between the monopoles. 
The distance of these monopoles are stabilized as
$g^2 Q/2\pi \tilde m$. 
This bound state is made of Abelian monopoles, 
but we can construct a bound state of non-Abelian monopoles  
by considering a set of masses degenerate in 
both real and imaginary parts,  
instead of Eq.~(\ref{eq:additional-masses}).

\section{Conclusion and Discussion}\label{sc:conclusion}

In this paper, we have studied domain walls in Abelian and 
non-Abelian gauge theories with degenerate masses for Higgs fields. 
In the model with degenerate masses, 
discrete components of the 
vacua are not necessarily isolated points 
but have continuous flat directions. 
Then the domain walls interpolating between these vacua have 
normalizable as well as non-normalizable
zero-modes corresponding to the Nambu-Goldstone 
modes of the broken non-Abelian flavor symmetry. 
When spatial infinities have such a degeneracy 
the wall solutions possess non-normalizable mode 
whose wave functions extend to infinity. 
On the other hand, when such a degeneracy appears 
between two domain walls, the wall solutions possess 
normalizable wave functions spreading  
between those domain walls. 
The latter are called non-Abelian clouds 
and appear in the effective theory 
on the domain walls. 
In the effective theory, these 
non-Abelian clouds give the long-range forces between 
two walls. 
We have constructed domain walls with stabilized relative 
position, which were supported by the long-range forces. 
They have different properties from those of Q-walls in 
models with fully non-degenerate masses. 
The properties of the domain walls in the model with 
degenerate masses have been discussed by using the D-brane 
configurations. 
We have 
determined the K\"ahler potential of the effective 
theory of the walls in the generalized Shifman-Yung 
model and have found that the effective dynamics of coincident 
walls are described by the chiral Lagrangian. 
In addition, we also have found that they are described by the 
chiral Lagrangian with mass terms if we introduce complex 
mass parameters which break the non-Abelian flavor symmetry 
in the original theory. 
We have performed the electromagnetic duality 
transformations to the massless scalars on the $3+1$ 
dimensional world-volume of the walls. 
We have obtained 
the antisymmetric tensor field with non-Abelian symmetry 
by applying the dual transformation of Freedman and 
Townsend.
We have given a brief discussion on the application to 
the non-Abelian monopoles confined by non-Abelian vortices. 
The possibility of a monopole-monopole bound state has been 
pointed out.

\bigskip


We give several discussions here. 

We have obtained supersymmetric extension of the $U(N)$ chiral Lagrangian. 
It is obviously interesting to include higher derivative corrections to it.
It was partially done \cite{Eto:2005cc} to obtain a four derivative term, 
which turned out to be the Skyrme term. 
Duality between Nambu-Goto type action and 
tensor gauge theory with higher derivative terms 
was discussed in \cite{Clark:2004jn}. 
So the dual tensor theory should be obtainable 
in the case with higher derivative terms.

Inclusion of a SUSY breaking term deserves to be studied. 
Since massless-ness of quasi-NG bosons is ensured only by supersymmetry 
they will acquire mass of the scale of SUSY breaking term.
There was large degeneracy of vacua so question is which vacuum is 
chosen by the SUSY breaking. 
This problem was studied in SUSY nonlinear sigma models \cite{Kotcheff:1988ji};
The answer is that there remain the vacua with the maximal unbroken symmetry. 
This implies that attractive force exist between 
(non-BPS) domain walls 
and then all the domain walls are compressed in the end.

Our work can be generalized to the case of 
domain wall networks 
which are 1/4 BPS states \cite{Eto:2005cp}. 
Non-Abelian clouds appear inside a domain wall loop there \cite{Eto:2007uc}.
In that case, a repulsive force caused by the charge given to
non-Abelian clouds is proportional to inverse of the area of the loop. 

Another interesting 1/4 BPS composite system is a system of  
vortices stretched between domain walls (called D-brane soliton) 
\cite{Gauntlett:2000de,Isozumi:2004vg}.
So far this system was studied in theories 
with non-degenerate Higgs masses, 
where domain walls are Abelian and possess 
only $U(1)$ internal moduli. 
Although the full exact solutions were already obtained 
\cite{Isozumi:2004vg} (in the strong gauge coupling limit), 
it is interesting to understand this configuration 
from the view point of the domain wall world volume. 
Vortex strings attached to domain walls can be 
regarded as sigma model lumps 
in the view point of the {\it total} domain wall moduli 
${\cal M}_{\rm total} \simeq G_{N_{\rm F},N_{\rm C}}
\simeq SU(N_{\rm F})/[SU(N_{\rm C})\times SU(N_{\rm F}-N_{\rm C})
\times U(1)]$, which can be  
constructed by patching all topological sectors together 
\cite{Isozumi:2004va}.\footnote{
The reason why we have to consider different topological 
sectors together is that 
the number of domain walls is reduced 
at the center of vortices \cite{Isozumi:2004va}. 
} 
The topological stability is ensured by 
$\pi_2 ({\cal M}_{\rm total}) \simeq {\bf Z}$. 
The other interpretation is that vortex strings 
can be regarded as  
{\it global} vortices of the $U(1)$ 
moduli of domain walls in the wall effective action. 
In this case, the topological stability is ensured by 
$\pi_1 [U(1)] \simeq {\bf Z}$.  
Domain walls with non-Abelian cloud possess non-Abelian moduli 
$U(N)$ as discussed in this paper. 
The total moduli space in the GSY model is 
${\cal M}_{\rm total} \simeq G_{2N,N} 
\simeq SU(2N)/[SU(N)\times SU(N)\times U(1)]$, 
so $\pi_2 ({\cal M}_{\rm total}) \simeq {\bf Z}$ 
as in the non-degenerate case. 
Sigma model lumps in this case however are more interesting. 
This is because the moduli space (with a fixed topological sector) 
is ${\cal M} \simeq T^* U(N)$ as we have seen in (\ref{eq:GSY-moduli}). 
This gives non-Abelian global vortices 
supported by $\pi_1[U(N)]\simeq {\bf Z}$
which are expected to form in the chiral phase transition 
\cite{Balachandran:2002je}.

\section*{Acknowledgments}
We would like to thank Youichi Isozumi for a collaboration 
at the early stage of this work.
ME, MN and KO would like to thank David Tong for a collaboration 
in \cite{Eto:2005cc}, 
Eric Weinberg and Kimyeong Lee 
for a fruitful discussion on non-Abelian clouds,  
and Korea Institute of Advanced Study (KIAS) for their hospitality. 
This work is supported in part by Grant-in-Aid for 
Scientific Research from the Ministry of Education, 
Culture, Sports, Science and Technology, Japan No.17540237
and No.18204024 (N.S.). 
The work of T.F.~is supported by the Research Fellowships 
of the Japan Society for the Promotion of Science for 
Young Scientists.
The work of M.E.~and K.O.~is also supported by the Research 
Fellowships of the Japan Society for 
the Promotion of Science for Research Abroad.


\appendix

\section{Asymptotic behavior of the K\"ahler potential}\label{appendix:1}
First, we consider the K\"ahler potential for one domain 
wall in $\Nf=2$ case as a simplest example. 
Let us take the moduli matrix and mass parameter as 
\beq
H_0 = ( 1 ,\, \phi), \hs{10} M = {\rm diag} \, (m,\,0).
\eeq
The position of the wall is given by 
\beq
y_0 \equiv \frac{1}{m} \log |\phi|.
\eeq
If we define $\psi \equiv \log \Omega$, the master 
equation (\ref{eq:master}) for one domain wall is written 
as 
\beq
\p_y^2 \psi= 
g^2 c \big( 1 - (e^{2m y} + e^{2m y_0}) e^{-\psi} \big).
\label{eq:master-psi}
\eeq
The asymptotic behavior of 
the solution $\psi$ far away from the wall position 
$y_0$ is given by 
\beq
\psi ~\simeq~ \log (e^{2my} + e^{2my_0}) ~\simeq~ \left\{ 
\begin{array}{lcl}
 2my   & & \text{for}~ y \gg  y_0 \\
 2my_0 & & \text{for}~ y \ll y_0 \\
\end{array}
\right., 
\label{eq:asymptotic}
\eeq
with exponentially suppressed correction terms of order 
${\cal O}(e^{-my})$ or ${\cal O}(e^{-yg\sqrt{c}})$ 
\cite{Isozumi:2003uh}, \cite{Sakai:2005kz}. 
The density of the K\"ahler potential 
(\ref{eq:formal-Kahler-density0}) can be written in terms 
of $\psi$ as 
\beq
\mathcal K = c \psi + c (e^{2my} + e^{2my_0}) e^{-\psi} 
+ \frac{1}{2g^2} (\p_y \psi)^2.
\label{eq:k-density}
\eeq 
The counter terms $\mathcal K_{ct}(\phi)$ and 
$\bar{\mathcal K}_{ct}(\phi^\ast)$, which are holomorphic 
and anti-holomorphic with respect to the moduli parameter 
$\phi$, are determined from the asymptotic behavior 
(\ref{eq:asymptotic}) as 
\beq
\mathcal K_{ct}(\phi) + \bar{\mathcal K}_{ct}(\phi^\ast) 
= c \Big[ 2my \, \theta(y) 
+ ( \log \phi + \log \phi^\ast) \theta(-y) \Big] 
+ c + \frac{2m^2}{g^2} \theta(y),
\label{eq:k-density-ct}
\eeq
where $\theta(y)$ is the step function. 
The K\"ahler potential can be calculated by using the 
transformation property under the translation such 
that $\psi(y+y_0,y_0) = \psi(y,0) + 2 m y_0$. 
Then we obtain 
the asymptotic behavior of the K\"ahler 
potential of one wall for large values of $y_0$ as
\beq
K ~=~ \int_{-\infty}^{\infty} dy \left( \mathcal K 
- \mathcal K_{ct}(\phi) - \bar{\mathcal K}_{ct}(\phi^\ast) 
\right) ~=~ mc \, y_0^2 - \frac{2 m^2}{g^2} y_0 
+ const.. 
\eeq

Next, let us calculate the K\"ahler potential for the 
walls with degenerate masses discussed in 
Section \ref{section:3}. 
The function $f(\mu R)$, which have been defined in 
(\ref{eq:f}), is independent of the number of the 
flavors with degenerate masses, so that we can calculate 
the function $f(\mu R)$ in $\Nf=3$ case. 
The moduli matrix, mass parameters and the master equation 
are given by 
\beq
&H_0 = (1,\,\phi,\,1), \hs{15} M 
= {\rm diag}\,(m_1,\,0,\,m_2),& \\
&\p_y^2 \psi = g^2c \Big( 1 - (e^{2m_1y} 
+ |\phi|^2 + e^{-2m_2y}) e^{-\psi} \Big).& 
\eeq
The positions of the walls are related to the 
parameter $\phi$ as
\beq
y_1 ~=~ \frac{1}{m_1} \log |\phi|, 
\hs{10} y_2 ~=~ - \frac{1}{m_2} \log |\phi|,
\eeq
and the relative distance of the walls is given by 
$R = y_1 - y_2 = 2/\mu \log |\phi|$ with 
$\mu \equiv 2 m_1 m_2/(m_1 + m_2)$. 
First, let us consider the 
asymptotic behavior of 
the K\"ahler potential for sufficiently large $R$. 
The solution of this master equation for sufficiently 
large $R$ is given by 
\beq
\psi \approx \psi_1 + \psi_2 - \mu R,
\label{eq:sol-asy}
\eeq
where $\psi_1$ and $\psi_2$ is the solution of the 
master equation for one wall (\ref{eq:master-psi}) 
with replacements 
$(y,\,y_0,\,m) \rightarrow (y,\,y_1,\,m_1)$ and 
$(y,\,y_0,\,m) \rightarrow (-y,\,-y_2,\,m_2)$ respectively. 
The correction to the solution (\ref{eq:sol-asy}) is 
exponentially small for sufficiently large $R$. 
For the solution of the master equation $\psi$, 
the density of the K\"ahler potential is written as 
\beq
\mathcal K &=& c \psi + c (e^{2m_1y} + e^{\mu R} 
+ e^{-2m_2y}) e^{-\psi} + \frac{1}{2g^2} (\p_y \psi)^2 
\notag \\
&\approx& \mathcal K_1 + \mathcal K_2 - c \mu R - c,
\eeq 
where we have used the fact that 
$\p_y \psi_1 \p_y \psi_2$ is exponentially small for 
large $R$. The counter terms are chosen to be 
\beq
\mathcal K_{ct} &=& 2 c \, y \Big[ m_1 \theta(y) - m_2 
\theta(-y) \Big] + c + \frac{2}{g^2} 
\Big[ (m_1)^2 \theta(y) + (m_2)^2 \theta(-y) \Big], \notag \\
&=& (\mathcal K_{ct})_1 + (\bar{\mathcal K}_{ct})_1 
+ (\mathcal K_{ct})_2 + (\bar{\mathcal K}_{ct})_2 
- c \mu R - c. \phantom{\frac12}
\eeq
Here the quantities with subscript 1,\,2 are given 
by the corresponding quantities (\ref{eq:k-density}) 
and (\ref{eq:k-density-ct}) with the replacements 
$(y,\,y_0,\,m) \rightarrow (y,\,y_1,\,m_1)$ and 
$(y,\,y_0,\,m) \rightarrow (-y,\,-y_2,\,m_2)$ respectively.
Then we find the asymptotic K\"ahler potential for large 
$R$ as 
\beq
K ~=~ \int_{-\infty}^{\infty} dy \left( \mathcal K - 
\mathcal K_{ct} \right) ~\approx~ K_1 + K_2 ~=~ 
\frac{c\mu}{2} R^2 - \frac{m_1+m_2}{g^2} \mu R + const..
\eeq
The correction to this K\"ahler potential is exponentially 
small for large $R$.
Next, let us consider asymptotic behavior for 
small 
$|\phi|^2 = e^{\mu R}$. 
The K\"ahler potential for sufficiently small $|\phi|$ 
can be easily obtained by assuming that the moduli space 
is smooth and $\phi$ is a good coordinate of the moduli 
space at $|\phi|=0$. Then the metric of the moduli space 
in terms of the coordinate $\phi$ can be expanded as 
\beq
g(|\phi|^2) ~\equiv~ \frac{\p^2 K}{\p \phi \p \phi^\ast} 
~=~ A + \mathcal O(|\phi|^2).
\eeq 
Here the constant term $A$ cannot be zero since $\phi$ is a 
good coordinate 
at $|\phi|=0$. 
Therefore the K\"ahler potential for small $|\phi|$ is given by
\beq
f(\mu R) ~=~ K(|\phi|^2) ~=~ A |\phi|^2 
+ \mathcal O(|\phi|^4) ~=~ A e^{\mu R} + \mathcal O(e^{2 \mu R}).
\eeq
Here we have ignored constant terms which do not 
contribute to the K\"ahler metric.

\section{General K\"ahler potential determined by symmetry}
\label{appendix:2}

In this subsection we construct the most 
general K\"ahler potential 
compatible with the symmetry (\ref{eq:isometry}) 
in the 
spirit 
of the method of nonlinear realization.
If we define
\beq
 X \equiv e^{\phi} e^{\phi^\dagger},
\eeq
it transforms as
\beq
 X \to g_{\rm L} X g_{\rm L}^\dagger.
\eeq
Then the most general K\"ahler potential 
invariant under the symmetry (\ref{eq:isometry}) is 
given using an arbitrary function $F$ of $N-1$ variables: 
\beq
 K = F ({\rm Tr} X, {\rm Tr} X^2, \cdots, {\rm Tr} X^{N-1}). 
  \label{eq:most-general-Kahler}
\eeq
Traces of higher order of $X$'s are not independent because of 
the Cayley-Hamilton theorem of $N$ by $N$ matrices $A$: 
$A^N - {\rm Tr} (A) A^{N-1} \cdots \pm (\det A) {\bf 1}_N =0$.
The K\"ahler potential (\ref{eq:most-general-Kahler})
was obtained by Shore long time ago \cite{Kotcheff:1988ji}.
The target space of this nonlinear sigma model is 
the complexification of $U(N)$: 
$GL(N,{\bf C}) = U(N)^{\rm C} \simeq T^* U(N)$.
By construction the isometry is not 
the transitive group 
${\bf C}^* \times SL(N,{\bf C})_{\rm L} \times 
SL(N,{\bf C})_{\rm R}$. 
The metric is invariant under 
its real subgroup (\ref{eq:isometry}) 
generated by a real form of the complex Lie algebra.
This always occurs if one constructs effective Lagrangian 
of massless particles when
a global symmetry is spontaneously 
broken with preserving supersymmetry \cite{Bando:1983ab}.
It was shown in \cite{Kotcheff:1988ji} 
that by setting quasi-NG modes zero 
the Lagrangian reduces at the most symmetric points 
(where $\phi$ is proportional to the unit matrix)
to the chiral Lagrangian of $U(N)$
${\cal L} = \1{2} f_{\pi}^2 {\rm Tr} [(U^\dagger \del_{\mu} U)^2]$, 
with $f_{\pi}^2 $ a constant determined by 
derivative of $F$.  
However at generic points symmetry is further broken and 
more Nambu-Goldstone bosons 
appear. 
It is known that 
$G$-invariants which are not invariant under $G^{\bf C}$, 
namely the variables 
${\rm Tr} X, {\rm Tr} X^2, \cdots, {\rm Tr} X^{N-1}$
in (\ref{eq:most-general-Kahler}), 
parametrize quasi-NG bosons 
at generic points in the moduli space \cite{Kotcheff:1988ji}.

Returning to our case of domain walls 
we have additional symmetry other than (\ref{eq:isometry}) 
so that we can further restrict the form of the K\"ahler potential 
(\ref{eq:most-general-Kahler}).
It is the translational symmetry of space-time broken by 
the presence of the domain walls: 
\beq
 \phi \to \phi + \lambda {\bf 1}_N , 
\quad 
X \to X e^{\lam + \lam^*}.
  \label{eq:translation}
\eeq
Interestingly this can be understood as the imaginary part of 
$e^{i \alpha} \in U(1)_{\rm A}$ in (\ref{eq:isometry}). 
The K\"ahler potential (\ref{eq:most-general-Kahler})
is reduced to
\beq
 K = c_1 \, {\rm Tr} [(\log X)^2] 
+ \tilde F\left[{({\rm Tr} X)^2 \over {\rm Tr} (X^2)},
 {({\rm Tr} X)^3 \over {\rm Tr} (X^3)}, 
{({\rm Tr} X) ({\rm Tr} X^2) \over {\rm Tr} (X^3)}, 
 \cdots \right] .
 \label{eq:kahler+}
\eeq
Here $\tilde F$ is 
an 
arbitrary function of variables 
with zero weight of $X$.  
The first term is invariant up to 
the K\"ahler transformation under 
the translational symmetry (\ref{eq:translation}),  
and the second term is strictly invariant under it. 
The first term is the K\"ahler potential 
(\ref{eq:compact-kahler}) 
with the identification of the overall constant $c_1 =c/4m$, 
and the second term describes the deformation 
of the metric along the non-compact directions 
with preserving the isometry \cite{Kotcheff:1988ji}.

\newcommand{\J}[4]{{\sl #1} {\bf #2} (#3) #4}
\newcommand{\andJ}[3]{{\bf #1} (#2) #3}
\newcommand{\AP}{Ann.\ Phys.\ (N.Y.)}
\newcommand{\MPL}{Mod.\ Phys.\ Lett.}
\newcommand{\NP}{Nucl.\ Phys.}
\newcommand{\PL}{Phys.\ Lett.}
\newcommand{\PR}{ Phys.\ Rev.}
\newcommand{\PRL}{Phys.\ Rev.\ Lett.}
\newcommand{\PTP}{Prog.\ Theor.\ Phys.}
\newcommand{\hep}[1]{{\tt hep-th/{#1}}}


\begin{thebibliography}{10}



\bibitem{Manton:1981mp}
N.~S.~Manton,
Phys.\ Lett.\ B {\bf 110}, 54 (1982); 
  N.~S.~Manton and P.~Sutcliffe,
  ``Topological solitons,''
{\it  Cambridge, UK: Univ. Pr. (2004)}.

\bibitem{Coleman:1969sm}
  S.~R.~Coleman, J.~Wess and B.~Zumino,
  Phys.\ Rev.\  {\bf 177}, 2239 (1969); 
  C.~G.~.~Callan, S.~R.~Coleman, J.~Wess and B.~Zumino,
  Phys.\ Rev.\  {\bf 177}, 2247 (1969).

\bibitem{Belavin:1975fg}
  A.~A.~Belavin, A.~M.~Polyakov, A.~S.~Schwarz and Yu.~S.~Tyupkin,
  Phys.\ Lett.\  B {\bf 59}, 85 (1975).

\bibitem{Dorey:2002ik}
  N.~Dorey, T.~J.~Hollowood, V.~V.~Khoze and M.~P.~Mattis,
  Phys.\ Rept.\  {\bf 371}, 231 (2002)
  [arXiv:hep-th/0206063].

\bibitem{Nekrasov:1998ss}
  N.~Nekrasov and A.~S.~Schwarz,
  Commun.\ Math.\ Phys.\  {\bf 198}, 689 (1998)
  [arXiv:hep-th/9802068].

\bibitem{Hanany:2003hp}
  A.~Hanany and D.~Tong,
  JHEP {\bf 0307}, 037 (2003)
  [arXiv:hep-th/0306150].

\bibitem{Auzzi:2003fs}
  R.~Auzzi, S.~Bolognesi, J.~Evslin, K.~Konishi and A.~Yung,
  Nucl.\ Phys.\ B {\bf 673}, 187 (2003)
  [arXiv:hep-th/0307287].

\bibitem{Abrikosov:1956sx}
  A.~A.~Abrikosov,
  Sov.\ Phys.\ JETP {\bf 5}, 1174 (1957)
  [Zh.\ Eksp.\ Teor.\ Fiz.\  {\bf 32}, 1442 (1957)];
  H.~B.~Nielsen and P.~Olesen,
  Nucl.\ Phys.\  B {\bf 61}, 45 (1973).

\bibitem{Eto:2005yh}
  M.~Eto, Y.~Isozumi, M.~Nitta, K.~Ohashi and N.~Sakai,
  %
  Phys.\ Rev.\ Lett.\  {\bf 96}, 161601 (2006)
  [arXiv:hep-th/0511088].

\bibitem{Tong:2005un}
  D.~Tong,
  ``TASI lectures on solitons,''
  arXiv:hep-th/0509216.

\bibitem{Eto:2006pg}
  M.~Eto, Y.~Isozumi, M.~Nitta, K.~Ohashi and N.~Sakai,
  J.\ Phys.\ A {\bf 39}, R315 (2006)
  [arXiv:hep-th/0602170].

\bibitem{Shifman:2007ce}
  M.~Shifman and A.~Yung,
  ``Supersymmetric Solitons and How They Help Us Understand Non-Abelian   Gauge
  Theories,''
  arXiv:hep-th/0703267.

\bibitem{Hanany:2004ea}
  A.~Hanany and D.~Tong,
  JHEP {\bf 0404}, 066 (2004)
  [arXiv:hep-th/0403158].

\bibitem{Eto:2004rz}
  M.~Eto, Y.~Isozumi, M.~Nitta, K.~Ohashi and N.~Sakai,
  %
  Phys.\ Rev.\ D {\bf 72}, 025011 (2005)
  [arXiv:hep-th/0412048].


\bibitem{Hanany:2005bq}
  A.~Hanany and D.~Tong,
  Commun.\ Math.\ Phys.\  {\bf 266}, 647 (2006)
  [arXiv:hep-th/0507140].

\bibitem{'t Hooft:1974qc}
  G.~'t Hooft,
  Nucl.\ Phys.\  B {\bf 79}, 276 (1974); 
  A.~M.~Polyakov,
  JETP Lett.\  {\bf 20}, 194 (1974)
  [Pisma Zh.\ Eksp.\ Teor.\ Fiz.\  {\bf 20}, 430 (1974)].



\bibitem{Abraham:1992vb}
E.~R.~C.~Abraham and P.~K.~Townsend,
Phys.\ Lett.\ B {\bf 291}, 85 (1992); 
Phys.\ Lett.\ B {\bf 295}, 225 (1992).

\bibitem{U(1)walls} 
  J.~P.~Gauntlett, D.~Tong and P.~K.~Townsend,
  Phys.\ Rev.\ D {\bf 64}, 025010 (2001)
  [arXiv:hep-th/0012178];
  D.~Tong,
  Phys.\ Rev.\ D {\bf 66}, 025013 (2002)
  [arXiv:hep-th/0202012];
  JHEP {\bf 0304}, 031 (2003)
  [arXiv:hep-th/0303151];
               K.~S.~M.~Lee, 
               Phys.\ Rev.\ D {\bf 67}, 045009 (2003) 
               [arXiv:hep-th/0211058];
M.~Arai, M.~Naganuma, M.~Nitta, and N.~Sakai, 
   Nucl.\ Phys.\ B {\bf 652},  35 (2003) [arXiv:hep-th/0211103]; 
``BPS Wall in N=2 SUSY Nonlinear Sigma Model with Eguchi-Hanson Manifold''
in Garden of Quanta - In honor of Hiroshi Ezawa, 
Eds. by J.~Arafune et al. 
(World Scientific Publishing Co. Pte. Ltd. Singapore, 2003) 
pp 299-325, [arXiv:hep-th/0302028]; 
M. Arai, E.~Ivanov and J.~Niederle, 
              Nucl.\ Phys.\ B {\bf 680},  23 (2004) 
              [arXiv:hep-th/0312037]. 
\bibitem{Isozumi:2003uh}
  Y.~Isozumi, K.~Ohashi and N.~Sakai,
  JHEP {\bf 0311}, 061 (2003)
  [arXiv:hep-th/0310130]; 
  JHEP {\bf 0311}, 060 (2003)
  [arXiv:hep-th/0310189].

\bibitem{Sakai:2005kz}
  N.~Sakai and Y.~Yang,
  Commun.\ Math.\ Phys.\  {\bf 267}, 783 (2006)
  [arXiv:hep-th/0505136].


\bibitem{Isozumi:2004jc}
  Y.~Isozumi, M.~Nitta, K.~Ohashi and N.~Sakai,
  %
  Phys.\ Rev.\ Lett.\  {\bf 93}, 161601 (2004)
  [arXiv:hep-th/0404198].

\bibitem{Isozumi:2004va}
  Y.~Isozumi, M.~Nitta, K.~Ohashi and N.~Sakai,
  %
  Phys.\ Rev.\ D {\bf 70}, 125014 (2004)
  [arXiv:hep-th/0405194].

\bibitem{Isozumi:2004vg}
  Y.~Isozumi, M.~Nitta, K.~Ohashi and N.~Sakai,
  %
  Phys.\ Rev.\ D {\bf 71}, 065018 (2005)
  [arXiv:hep-th/0405129].

\bibitem{Eto:2004vy}
  M.~Eto, Y.~Isozumi, M.~Nitta, K.~Ohashi, K.~Ohta and N.~Sakai,
  %
  Phys.\ Rev.\ D {\bf 71}, 125006 (2005)
  [arXiv:hep-th/0412024].

\bibitem{Eto:2005wf}
  M.~Eto, Y.~Isozumi, M.~Nitta, K.~Ohashi, K.~Ohta, N.~Sakai and Y.~Tachikawa,
  %
  Phys.\ Rev.\ D {\bf 71}, 105009 (2005)
  [arXiv:hep-th/0503033].



\bibitem{Shifman:2003uh}
  M.~Shifman and A.~Yung,
  %
  Phys.\ Rev.\ D {\bf 70}, 025013 (2004)
  [arXiv:hep-th/0312257].

\bibitem{Eto:2005cc}
  M.~Eto, M.~Nitta, K.~Ohashi and D.~Tong,
  %
  Phys.\ Rev.\ Lett.\  {\bf 95}, 252003 (2005)
  [arXiv:hep-th/0508130].


\bibitem{Tong:2003pz}
  D.~Tong,
  Phys.\ Rev.\  D {\bf 69}, 065003 (2004)
  [arXiv:hep-th/0307302].


\bibitem{Shifman:2004dr}
  M.~Shifman and A.~Yung,
  Phys.\ Rev.\  D {\bf 70}, 045004 (2004)
  [arXiv:hep-th/0403149].

\bibitem{NA-monopoles}
  P.~Goddard, J.~Nuyts and D.~I.~Olive,
  Nucl.\ Phys.\  B {\bf 125}, 1 (1977);
  F.~A.~Bais,
  Phys.\ Rev.\  D {\bf 18}, 1206 (1978); 
  E.~J.~Weinberg,
  Nucl.\ Phys.\  B {\bf 167}, 500 (1980);
  Phys.\ Lett.\  B {\bf 119}, 151 (1982);
  Nucl.\ Phys.\  B {\bf 203}, 445 (1982);
  A.~P.~Balachandran, G.~Marmo, N.~Mukunda, J.~S.~Nilsson, E.~C.~G.~Sudarshan and F.~Zaccaria,
  Phys.\ Rev.\ Lett.\  {\bf 50}, 1553 (1983);
  Phys.\ Rev.\  D {\bf 29}, 2919 (1984);
  Phys.\ Rev.\  D {\bf 29}, 2936 (1984);
  M.~K.~Murray,
  Commun.\ Math.\ Phys.\  {\bf 96}, 539 (1984).
  K.~M.~Lee, E.~J.~Weinberg and P.~Yi,
  Phys.\ Rev.\  D {\bf 54}, 6351 (1996)
  [arXiv:hep-th/9605229];
  C.~h.~Lu,
  Phys.\ Rev.\  D {\bf 58}, 125010 (1998)
  [arXiv:hep-th/9806237];
  E.~J.~Weinberg,
  ``Massive monopoles and massless monopole clouds,''
in the proceedings of International Workshop on Mathematical and Physical Aspects of Nonlinear Field Theories, Seoul, Korea, 23-24 Feb 1998 
  [arXiv:hep-th/9908097];
  X.~Chen and E.~J.~Weinberg,
  Phys.\ Rev.\  D {\bf 64}, 065010 (2001)
  [arXiv:hep-th/0105211];
  X.~Chen, H.~d.~Guo and E.~J.~Weinberg,
  Phys.\ Rev.\  D {\bf 64}, 125004 (2001)
  [arXiv:hep-th/0108029];
  C.~J.~Houghton and E.~J.~Weinberg,
  Phys.\ Rev.\  D {\bf 66}, 125002 (2002)
  [arXiv:hep-th/0207141].

\bibitem{Weinberg:2006rq}
  E.~J.~Weinberg and P.~Yi,
  Phys.\ Rept.\  {\bf 438}, 65 (2007)
  [arXiv:hep-th/0609055].




\bibitem{Dorey:1996jh}
  N.~Dorey, C.~Fraser, T.~J.~Hollowood and M.~A.~C.~Kneipp,
  Phys.\ Lett.\  B {\bf 383}, 422 (1996)
  [arXiv:hep-th/9605069];
  arXiv:hep-th/9512116;
  M.~J.~Strassler,
  JHEP {\bf 9809}, 017 (1998)
  [arXiv:hep-th/9709081];
  F.~A.~Bais and B.~J.~Schroers,
  Nucl.\ Phys.\  B {\bf 512}, 250 (1998)
  [arXiv:hep-th/9708004];
  B.~J.~Schroers and F.~A.~Bais,
  Nucl.\ Phys.\  B {\bf 535}, 197 (1998)
  [arXiv:hep-th/9805163].



\bibitem{Auzzi:2004if}
  R.~Auzzi, S.~Bolognesi, J.~Evslin, K.~Konishi and H.~Murayama,
  Nucl.\ Phys.\ B {\bf 701}, 207 (2004)
  [arXiv:hep-th/0405070]; 
  R.~Auzzi, S.~Bolognesi, J.~Evslin and K.~Konishi,
  Nucl.\ Phys.\  B {\bf 686}, 119 (2004)
  [arXiv:hep-th/0312233];
  K.~Konishi,
  ``The magnetic monopoles seventy-five years later,''
  arXiv:hep-th/0702102.

\bibitem{Eto:2006uw}
  M.~Eto, Y.~Isozumi, M.~Nitta, K.~Ohashi and N.~Sakai,
  %
  Phys.\ Rev.\ D {\bf 73}, 125008 (2006)
  [arXiv:hep-th/0602289].

%
\bibitem{Bando:1983ab}
  M.~Bando, T.~Kuramoto, T.~Maskawa and S.~Uehara,
  Phys.\ Lett.\  B {\bf 138}, 94 (1984);
  Prog.\ Theor.\ Phys.\  {\bf 72}, 313 (1984);
  Prog.\ Theor.\ Phys.\  {\bf 72}, 1207 (1984);
  K.~Higashijima, M.~Nitta, K.~Ohta and N.~Ohta,
  Prog.\ Theor.\ Phys.\  {\bf 98}, 1165 (1997)
  [arXiv:hep-th/9706219]; 
  K.~Higashijima and M.~Nitta,
  Prog.\ Theor.\ Phys.\  {\bf 103}, 635 (2000)
  [arXiv:hep-th/9911139];
  Prog.\ Theor.\ Phys.\  {\bf 103}, 833 (2000)
  [arXiv:hep-th/9911225];
  M.~Nitta,
  Nucl.\ Phys.\  B {\bf 711}, 133 (2005)
  [arXiv:hep-th/0312025].

\bibitem{Kotcheff:1988ji}
  A.~C.~W.~Kotcheff and G.~M.~Shore,
  Int.\ J.\ Mod.\ Phys.\  A {\bf 4}, 4391 (1989); 
  Nucl.\ Phys.\  B {\bf 301}, 267 (1988).
  G.~M.~Shore,
  Nucl.\ Phys.\  B {\bf 320}, 202 (1989);
  Nucl.\ Phys.\  B {\bf 334}, 172 (1990); 
  M.~Nitta,
  Int.\ J.\ Mod.\ Phys.\  A {\bf 14}, 2397 (1999)
  [arXiv:hep-th/9805038].

\bibitem{Buchmuller:1982xn}
  W.~Buchmuller, S.~T.~Love, R.~D.~Peccei and T.~Yanagida,
  Phys.\ Lett.\  B {\bf 115}, 233 (1982).

\bibitem{Clark:1988gx}
  T.~E.~Clark, C.~H.~Lee and S.~T.~Love,
  Mod.\ Phys.\ Lett.\  A {\bf 4}, 1343 (1989);
  K.~Furuta, T.~Inami, H.~Nakajima and M.~Nitta,
  Prog.\ Theor.\ Phys.\  {\bf 106}, 851 (2001)
  [arXiv:hep-th/0106183].

\bibitem{Freedman:1980us}
  D.~Z.~Freedman and P.~K.~Townsend,
  %
  Nucl.\ Phys.\ B {\bf 177}, 282 (1981).





  \bibitem{brane-world}     
    P.~Horava and E.~Witten, 
     Nucl.\ Phys.\ {\bf B460}, 506 (1996) [arXiv:hep-th/9510209]. 
    N.~Arkani-Hamed, S.~Dimopoulos and G.~Dvali, 
             Phys.\ Lett.\ {\bf B429}, 263  (1998) 
             [arXiv:hep-ph/9803315]; 
             I.~Antoniadis, N.~Arkani-Hamed, S.~Dimopoulos 
             and G.~Dvali, 
             Phys.\ Lett.\ {\bf B436}, 257  (1998) 
             [arXiv:hep-ph/9804398]; 
    L.~Randall and R.~Sundrum, 
             Phys.\ Rev.\ Lett.\ 
             {\bf 83}, 3370 (1999)  [arXiv:hep-ph/9905221]; 
             Phys.\ Rev.\ Lett.\ {\bf 83}, 4690  (1999) 
             [arXiv:hep-th/9906064].



\bibitem{Fayet:1974jb}
  P.~Fayet and J.~Iliopoulos,
  Phys.\ Lett.\ B {\bf 51} (1974) 461.

\bibitem{Lindstrom:1983rt}
  U.~Lindstrom and M.~Rocek,
  Nucl.\ Phys.\  B {\bf 222}, 285 (1983).

\bibitem{Arai:2003tc}
  M.~Arai, M.~Nitta and N.~Sakai,
  Prog.\ Theor.\ Phys.\  {\bf 113}, 657 (2005)
  [arXiv:hep-th/0307274]; 
  Phys.\ Atom.\ Nucl.\  {\bf 68}, 1634 (2005)
  [Yad.\ Fiz.\  {\bf 68}, 1698 (2005)]
  [arXiv:hep-th/0401102].

\bibitem{Inami:2006wr}
  T.~Inami, S.~Minakami and M.~Nitta,
  Nucl.\ Phys.\  B {\bf 752}, 391 (2006)
  [arXiv:hep-th/0605064].

\bibitem{Eto:2006mz}
  M.~Eto, T.~Fujimori, Y.~Isozumi, M.~Nitta, K.~Ohashi, K.~Ohta and N.~Sakai,
  Phys.\ Rev.\  D {\bf 73}, 085008 (2006)
  [arXiv:hep-th/0601181];
  M.~Eto, K.~Konishi, G.~Marmorini, M.~Nitta, K.~Ohashi, W.~Vinci and N.~Yokoi,
  Phys.\ Rev.\  D {\bf 74}, 065021 (2006)
  [arXiv:hep-th/0607070];
  M.~Eto, K.~Hashimoto, G.~Marmorini, M.~Nitta, K.~Ohashi and W.~Vinci,
  Phys.\ Rev.\ Lett.\  {\bf 98}, 091602 (2007)
  [arXiv:hep-th/0609214];
  M.~Eto, L.~Ferretti, K.~Konishi, G.~Marmorini, M.~Nitta, K.~Ohashi, W.~Vinci and N.~Yokoi,
  Nucl.\ Phys.\  B {\bf 780}, 161 (2007)
  [arXiv:hep-th/0611313];
   M.~Eto, J.~Evslin, K.~Konishi, G.~Marmorini, M.~Nitta, K.~Ohashi, W.~Vinci and N.~Yokoi,  
  Phys.\ Rev.\  D {\bf 76}, 105002 (2007)
  [arXiv:0704.2218 [hep-th]];
  M.~Eto, T.~Fujimori, M.~Nitta, K.~Ohashi, K.~Ohta and N.~Sakai,
  Nucl.\ Phys.\  B {\bf 788}, 120 (2008)
  [arXiv:hep-th/0703197]; 
  M.~Eto, T.~Fujimori, S.~B.~Gudnason, K.~Konishi, M.~Nitta, K.~Ohashi and W.~Vinci,
  arXiv:0802.1020 [hep-th].



%
\bibitem{proceedings}
Y.~Isozumi, M.~Nitta, K.~Ohashi and N.~Sakai, 
Proceedings of 12th International Conference on 
Supersymmetry and Unification of Fundamental Interactions 
(SUSY 04), Tsukuba, Japan, 17-23 Jun 2004,  
edited by K. Hagiwara {\it et al.} (KEK, 2004) p.1 - p.16 
 [arXiv:hep-th/0409110]; 
  ``Walls and vortices in supersymmetric non-Abelian gauge theories,''
  to appear in the proceedings of ``NathFest'' at PASCOS 
conference, 
Northeastern University, Boston, Ma, August 2004  
  [arXiv:hep-th/0410150];
  M.~Eto, Y.~Isozumi, M.~Nitta, K.~Ohashi and N.~Sakai,
  ``Solitons in supersymmetric gauge theories,''
AIP Conf. Proc. {\bf 805}, 266 (2005) [arXiv:hep-th/0508017];
  M.~Eto, Y.~Isozumi, M.~Nitta, K.~Ohashi and N.~Sakai,
  ``Solitons in supersymmetric gauge theories: Moduli matrix approach,''
pages 58-71, in
``Continuous Advances in QCD 2006''C
(2007) World Scientific Pub. Singapore
Proceedings of the conference Continuous Advances in QCD 2006,
held at Univ. of Minnesota
May 11-May 14 2006,
[arXiv:hep-th/0607225]; 
  N.~Sakai, M.~Eto, Y.~Isozumi, M.~Nitta and K.~Ohashi,
  in PoS stringsLHC 2006:025,2006 [arXiv:hep-th/0703136];
  N.~Sakai, M.~Eto, T.~Fujimori, T.~Nagashima, M.~Nitta and K.~Ohashi,
  Submitted for the SUSY07 proceedings [arXiv:0710.0423 [hep-th]].





\bibitem{Lambert:1999ix}
  N.~D.~Lambert and D.~Tong,
  %
  Nucl.\ Phys.\ B {\bf 569}, 606 (2000)
  [arXiv:hep-th/9907098].


\bibitem{Eto:2005cp}
  M.~Eto, Y.~Isozumi, M.~Nitta, K.~Ohashi and N.~Sakai,
  %
  Phys.\ Rev.\ D {\bf 72}, 085004 (2005)
  [arXiv:hep-th/0506135];
  %
  Phys.\ Lett.\ B {\bf 632}, 384 (2006)
  [arXiv:hep-th/0508241];
  M.~Eto, Y.~Isozumi, M.~Nitta, K.~Ohashi, K.~Ohta and N.~Sakai,
  %
  AIP Conf.\ Proc.\  {\bf 805}, 354 (2005)
  [arXiv:hep-th/0509127];
  M.~Eto, T.~Fujimori, T.~Nagashima, M.~Nitta, K.~Ohashi and N.~Sakai
  Phys.\ Rev.\  D {\bf 75}, 045010 (2007)
  [arXiv:hep-th/0612003].

\bibitem{Eto:2007uc}
  M.~Eto, T.~Fujimori, T.~Nagashima, M.~Nitta, K.~Ohashi and N.~Sakai,
  Phys.\ Rev.\  D {\bf 76}, 125025 (2007)
  [arXiv:0707.3267 [hep-th]].


\bibitem{Lee:2005sv}
  K.~Lee and H.~U.~Yee,
  %
  Phys.\ Rev.\ D {\bf 72}, 065023 (2005)
  [arXiv:hep-th/0506256].

\bibitem{Eto:2005sw}
  M.~Eto, Y.~Isozumi, M.~Nitta and K.~Ohashi,
  Nucl.\ Phys.\  B {\bf 752}, 140 (2006)
  [arXiv:hep-th/0506257].

\bibitem{HananyWitten97}
  A.~Hanany and E.~Witten,
  Nucl.\ Phys.\  B {\bf 492}, 152 (1997)
  [arXiv:hep-th/9611230].



\bibitem{Higashijima:2001vk}
  K.~Higashijima, T.~Kimura and M.~Nitta,
  Nucl.\ Phys.\  B {\bf 623}, 133 (2002)
  [arXiv:hep-th/0108084];
  Annals Phys.\  {\bf 296}, 347 (2002)
  [arXiv:hep-th/0110216]; 
  Nucl.\ Phys.\  B {\bf 645}, 438 (2002)
  [arXiv:hep-th/0202064].

\bibitem{Eto:2004ii}
  M.~Eto, M.~Nitta and N.~Sakai,
  Nucl.\ Phys.\  B {\bf 701}, 247 (2004)
  [arXiv:hep-th/0405161].

\bibitem{Hashimoto:2007fa}
  K.~Hashimoto, T.~Hirayama and A.~Miwa,
  JHEP {\bf 0706}, 020 (2007)
  [arXiv:hep-th/0703024].

\bibitem{WB}
J.~Wess and J.~Bagger,
 {\sl Supersymmetry and Supergravity} 
(Princeton Univ. Press, Princeton, 1992).

\bibitem{Siegel:1979ai}
  W.~Siegel,
  Phys.\ Lett.\  B {\bf 85}, 333 (1979).

\bibitem{Gates:1983nr}
  S.~J.~Gates, M.~T.~Grisaru, M.~Rocek and W.~Siegel,
  ``Superspace, or one thousand and one lessons in supersymmetry,''
  Front.\ Phys.\  {\bf 58}, 1 (1983)
  [arXiv:hep-th/0108200].

\bibitem{Clark:2004jn}
  T.~E.~Clark, M.~Nitta and T.~ter Veldhuis,
  Phys.\ Rev.\  D {\bf 70}, 125011 (2004)
  [arXiv:hep-th/0409151].

\bibitem{Gauntlett:2000de}
  J.~P.~Gauntlett, R.~Portugues, D.~Tong and P.~K.~Townsend,
  Phys.\ Rev.\  D {\bf 63}, 085002 (2001)
  [arXiv:hep-th/0008221]; 
  M.~Shifman and A.~Yung,
  Phys.\ Rev.\  D {\bf 67}, 125007 (2003)
  [arXiv:hep-th/0212293].

\bibitem{Balachandran:2002je}
  A.~P.~Balachandran and S.~Digal,
  Phys.\ Rev.\  D {\bf 66}, 034018 (2002)
  [arXiv:hep-ph/0204262];
  M.~Nitta and N.~Shiiki,
  Phys.\ Lett.\  B {\bf 658}, 143 (2008)
  [arXiv:0708.4091 [hep-ph]];
  E.~Nakano, M.~Nitta and T.~Matsuura,
  arXiv:0708.4092 [hep-ph].




\end{thebibliography}
\end{document}